\documentclass[review]{elsarticle}

\usepackage{lineno,hyperref}
\modulolinenumbers[5]
\usepackage{amsmath}
\usepackage{amsfonts}
\usepackage{amssymb}
\usepackage{geometry}
\usepackage{times}
\usepackage{soul}
\usepackage{url}
\usepackage[utf8]{inputenc}
\usepackage{bbding}
\usepackage{graphicx}
\usepackage{amsmath}
\usepackage{booktabs}
\usepackage{lineno}
\usepackage{pifont}
\usepackage{graphicx}
\usepackage{subfigure}
\usepackage{caption}   % \ContinuedFloat宏包
\usepackage{multirow}
\usepackage{amsmath}
\usepackage{amsfonts}
\usepackage{color}
\usepackage{multirow}
\usepackage{booktabs}
\usepackage{enumerate}
\usepackage{float}
\usepackage{hyperref}       % hyperlinks
\usepackage{nicefrac}       % compact symbols for 1/2, etc.
\usepackage{times}
\usepackage{xcolor}
\usepackage{lineno,hyperref}
\usepackage{enumerate}
\usepackage{amsfonts}       % blackboard math symbols
\usepackage{tcolorbox}
\usepackage{diagbox}
\usepackage{algorithm}
\usepackage{algorithmicx}
\usepackage{algpseudocode}

\graphicspath{ {images/} }
\begin{document}

\begin{frontmatter}

\title{GAIL-PT: A Generic Intelligent Penetration Testing Framework with \\ Generative Adversarial Imitation Learning  }

\author[mymainaddress,mysecondaryaddress]{Jinyin Chen}
\ead{chenjinyin@zjut.edu.cn}

\author[mysecondaryaddress]{Shulong Hu}
\ead{2112003213@zjut.edu.cn}

\author[mysecondaryaddress]{Haibin Zheng}
\ead{haibinzheng320@gmail.com}

\author[mythirdaryaddress]{Changyou Xing\corref{mycorrespondingauthor}}
\cortext[mycorrespondingauthor]{Corresponding author}
\ead{changyouxing@126.com}

\author[mythirdaryaddress]{Guomin Zhang}
\ead{zhanggmwn@163.com}

\address[mymainaddress]{Institute of Cyberspace Security, Zhejiang University of Technology, Hangzhou, China}
\address[mysecondaryaddress]{College of Information Engineering, Zhejiang University of Technology, Hangzhou, China}
\address[mythirdaryaddress]{College of Command and Control Engineering, Army Engineering University of PLA, Nanjing, China}

\begin{abstract}
Penetration testing (PT) is an efficient network testing and vulnerability mining tool by simulating a hacker's attack for valuable information applied in some areas, e.g., the host's operating and database systems. Compared with manual PT, intelligent PT has become a dominating mainstream due to less time-consuming and lower labor costs. Among the already proposed intelligent PT methods, reinforcement learning (RL) based PT has achieved state-of-the-art (SOTA) performance. Unfortunately, RL-based PT is still challenged in real exploitation scenarios because the agent's action space is usually high-dimensional discrete, thus leading to algorithm convergence difficulty. Besides, most PT methods still rely on the decisions of security experts. Addressing the challenges, for the first time, we introduce expert knowledge to guide the agent to make better decisions in RL-based PT, and propose a \emph{\underline{G}enerative \underline{A}dversarial \underline{I}mitation \underline{L}earning based generic intelligent \underline{P}enetration \underline{T}esting framework}, denoted as GAIL-PT, to solve the problems of higher labor costs due to the involvement of security experts and high-dimensional discrete action space. Specifically,  first, we manually collect the state-action pairs to construct an expert knowledge base when the pre-trained RL / DRL model executes successful penetration testings. Second, we input the expert knowledge and the state-action pairs generated online by the different RL / DRL models into the discriminator of GAIL for training. At last, we apply the output reward of the discriminator to guide the agent to perform the action with a higher penetration success rate to improve PT's performance. Extensive experiments conducted on the real target host and simulated network scenarios show that GAIL-PT achieves SOTA penetration performance against DeepExploit in exploiting actual target Metasploitable2 and Q-learning in optimizing penetration path, not only in small-scale with or without honey-pot network environments, but also in the large-scale simulated network environment. The code of GAIL-PT is open-sourced at \textit{\url{https://github.com/Shulong98/GAIL-PT//}}.
\end{abstract}


\end{frontmatter}

\section{Introduction}
Penetration Testing (PT) is a process of simulating the attack behavior of malicious hackers, launching authorized attacks on computer systems and networks to discover any security vulnerabilities that can be exploited~\cite{2}\cite{1}. Numerous PT methods have been proposed for assessing the security of network systems, such as host operating systems \cite{os}, database systems \cite{ds}, network equipment systems \cite{firewall1} \cite{firewall2} and etc.. According to different testing techniques, PT methods are roughly categorized as manual PT, automatic PT and intelligent PT.

Manual PT depends on applying software exploits like running computer code to trigger a system vulnerabilities and taking further actions such as payloads that after a successful exploit, according to the security audit experts\cite{aduit}. Generally speaking, they take full advantage of some penetration tools (e.g., Nessus \cite{nessus}, Nexpose \cite{nexpose} ) combined with expert operations to accomplish the PT process. However, those tools rely on the professional knowledge of security experts to make decisions, so they usually can not update attack strategies accordingly when the environment changes. Additionally, with the complexity of systems grows, the task of manually assessing security will become quite a burden of labor-consuming and time-consuming. Consequently, the emergence of automatic PT methods alleviate this problem to a certain extent.

Existing approaches of the automatic PT include those mapping the results of vulnerability scanners to the corresponding penetration tools and those describing the PT process as an attack graph to solve the problem of the penetration path planning. The former method is integrated by vulnerability scanning, penetration attacking, payloads choosing and other modules into a framework to realize PT automatically, e.g., Metasploit \cite{3}\cite{metasploit}, Core Impact \cite{coreimpact}. However, some operations still require manual participation in the critical penetration steps, such as selecting proper exploit modules and payloads. In other words, it still relies on the attack rules carefully set in advance by programmers and security experts \cite{4}; thus, it cannot learn to deal with dynamic and uncertain penetration environments to develop new attack strategies. The latter method of automatic PT is mainly based on attack graphs \cite{graph}, by showing the attack sequence and the effect that an attacker may launch. In the graph, the attacker could use the relationship of vulnerabilities and exploits that are already grasp to choose an attack path. Nevertheless, since the attack graphs demands the complete system knowledge, it is difficult to apply to realistic dynamic penetration environments. To address the problem, Artificial Intelligence (AI) based PT methods come into being.

According to the AI techniques adopted for PT, AI-based intelligent PT methods can be mainly grouped into PT based on traditional Reinforcement Learning (RL) and Deep Reinforcement Learning (DRL) \cite{DRL}.Specifically, in PT based on traditional RL, such as modeling the attack process as a Partially Observable Markov Decision Process (POMDP) \cite{POMDP2}\cite{POMDP1}, the attacker gradually observes and models the computer configuration as the attack progresses. It can be well-aimed at a single host in practice, but due to POMDP's high computational complexity, it is not scalable for large networks \cite{POMDP3}. Besides, the Q-learning algorithm is also applied to the Capture The Flag (CTF) competition \cite{CTF} and the penetration path optimization \cite{pathfind}. However, these methods are also challenged in real-world scenarios since they are hard to involve in large-scale networks. Besides, the RL agent needs to comprehend the complete network topology in advance, which is a strong assumption in the real environment. Furthermore, intelligent PT based on DRL uses the function approximator as a neural network \cite{DRL} to conduct PT. It is easy to encounter the high-dimensional discrete of the action space when applied to the actual scenario and complex network system, which makes DRL model tough to train and converge in the PT process. Taking the value-based Deep Q-Network (DQN) algorithm \cite{DQN} as an example, the limitation of using DQN for automated PT lies in the complexity of the action space of different penetration scenarios. DQN evaluates all output actions' values, and selects the largest value as the best option in the entire action space. If the action space is large, we may encounter the problem that multiple actions have the same value \cite{growingactionspace}. PT is similar; the state and action space exponentially expand with the increasing hosts in different subnets. Consequently, applying the existing DRL algorithms to automate PT is difficult and unstable. Although in a relatively small network scene, it still faces the challenge of the state and action space explosion with hundreds or thousands of dimensions \cite{explode}.
 
In summary, the existing PT methods are still challenged in three aspects: (1) Security experts play a significant role in the PT process, but over-reliance on security experts' manual operations and decisions will also increase the labor cost of PT;
(2) When using the DRL algorithm for automating PT, it could encounter the problem of large state space and high-dimensional discrete action space, thus leading to convergence difficulty of PT training; (3) Most intelligent PT methods are verified in the virtual network environment instead of real penetration scenarios.

To address these challenges, for the first time, we introduce automatically-collected expert knowledge for agents to perform autonomous PT,  proposing a general intelligent PT framework that incorporates Generative Adversarial Imitation Learning, denoted as GAIL-PT.We combine GAIL with RL / DRL to automate PT process, by modeling the penetration attacker as a agent. Specifically, due to imitation learning has achieved a great success for speeding up the convergence rate of reinforcement learning with learned strategy in prior, we adopt GAIL for accumulated expert knowledge for RL / DRL based PT. Then, the agent continuously interacts with the penetration environment through the guidance of expert knowledge, and the accumulated penetration experience will assist the agent in formulating new attack strategies with a higher penetration success rate adaptive to the dynamic environments. At last, for both  small-scale and large-scale network scenes with high-dimensional discrete action space, GAIL-PT shows state-of-the-art (SOTA) penetration performance compared with baselines and also can be verified in practical scenarios.

The main contributions of this paper are as follows:
\begin{itemize}
\item Aiming at the problem that most PT methods over-reliance on security experts' manual operations and decisions, we construct an expert knowledge base and store state-action pairs as experiences in different penetration scenarios so that the PT experts can take part in the decision-making in a lower-cost way.

\item To solve the convergence difficulty problem of PT training and improve the penetration performance, we apply the GAIL network to PT based on RL or DRL for the first time. With the guidance of the expert knowledge, the agent is trained through the GAIL network to perform the action infinitely closed to the expert knowledge base, making the training process more stable and efficient for PT.

\item We conduct extensive experiments not only in real target Metasploitable2 and small-scale networks with or without honeypot, but also for large-scale networks. The experiment results indicate that GAIL-PT shows the SOTA penetration performance in actual or simulated network scenes. In addition, more experiments also verified that the proposed GAIL-PT is a general leading framework suitable for different reinforcement learning methods, i.e., DRL, RL. 

\end{itemize}
 The organization of this paper is as follows. Section~\ref{RWs} provides a overview of penetration testing, penetration testing based on reinforcement learning and imitation learning. Section~\ref{Preliminaries} supplies an overview of the reinforcement learning models we applied. Section~\ref{Method} exhibits the outline of our method and demonstrates its details. Section~\ref{Exp} describes the experimental setup, then evaluates and discusses the results. Finally, in Section~\ref{Conclusion and future work} the main findings and limitations of our work are summarized.

\section{Related Work\label{RWs}}
In this section, we mainly present penetration testing, the current research status of penetration testing based on reinforcement learning and mainstream imitation learning methods.

\subsection{Penetration testing}
PT is a security exercise designed to evaluate the system's overall security by authorizing simulated cyberattacks on the computer system. Manual PT relies on the use of penetration tools. For instance, Nessus \cite{nessus} and Nmap \cite{nmap} use the target list to perform network vulnerability testing one by one without checking for real-time environment changes. They neither make any decisions automatically, nor work independently. As one of the best performing vulnerability exploits, Core Impact \cite{coreimpact}, establishes an attack plan automatically by security experts. However, its limitation lies in generating attack plans on the target environment before conducting PT. Consequently, Core Impact cannot be dynamically adaptive to the environment either.  Another prevalent PT tool is Rapid 7 Nexpose \cite{nexpose}. It uses the vulnerability development, and penetration tool Metasploit \cite{metasploit} for vulnerability scanning and penetration. Although it can achieve flexibility by distributing multiple network scan engines, Nexpose is unable to learn attack policies. In short, most well-known commercial penetration testing tools show lots of advantages but highly rely on the expertise and decision-making of PT experts. These tools suffer from the problem of poor adaptability, that is, once the environment changes, the attack strategy cannot be updated accordingly.

In addition to manual PT methods, automated PT methods have been further explored. Numerous automated PT methods define PT as a path planning problem for constructing attack graphs \cite{BID6}. Cynthia and Swiler \cite{graph} firstly proposed a method based on the attack graph to analyze the system's vulnerability. The system divides the common attack database into atomic steps, distinct network configuration, topology details and attacker configuration files. The nodes and chain edges in the attack graph describe the different attack stages. The probability of success will be allocated to the chain edge, so different attack graph algorithms are used to identify the attack path with the highest probability of success. Based on the previous attack graph method, Kyle et al. \cite{BID8} created the NetSPA attack graphics system; its innovation lies in allowing network defenders to assess security threats to select complementary strategies. NetSPA analyzes multiple targets in minutes using firewall rules and vulnerability scanning, drastically reducing attack graph building time. Considering the attack path's reliability, Xue Qiu et al. \cite{BID9} proposed an algorithm for automatically generating attack graphs using Common Vulnerability Scoring System (CVSS) \cite{cvss} for the first time to improve the trustworthiness of attack paths by eventually optimizing the network topology. As a result, once an attack graph is generated, the network topology is also optimized. However, most of these methods could simply output corresponding action sequences to handle static environments. In other words, they can go back and provide corresponding steps or guidelines for PT,  but cannot perform PT dynamically and interactively in real PT scenarios. In summary, we concluded some details of the introduced manual and automatic PT tools as shown in Table~\ref{tabel:PT drawbacks}.

\begin{table}[!htp]
     \newcommand{\tabincell}[2]{\begin{tabular}{@{}#1@{}}#2\end{tabular}}
     \footnotesize
     \centering
     \caption{The details of manual and automatic PT tools}
     \label{tabel:PT drawbacks}
     \resizebox{1.00\linewidth}{!}
     {\begin{tabular}{cccccc}
               \toprule \hline
               \textbf{Category} & \textbf{Tool}  &\textbf{Benefits} &\textbf{Drawbacks} &\textbf{Scenes} &\textbf{Integration}\\
               \hline
\multirow{3}{*}{Manual PT}
     &{Nessus}\cite{nessus} & \multirow{3}*{{Vulnerability scanning precisely }} & \multirow{3}*{{Cannot learn attack strategies}}  &\multirow{3}*{{Single server \& Network topology}}  &\multirow{3}*{\tabincell{c}{Low}} \\
\cline{2-2}
&{Nmap}\cite{nmap}&&&&  \\
               \cline{2-2}
&{Nexpose}\cite{nexpose}&&&     &  \\
          \hline
               \multirow{3}*{Automatic PT}
          & {Metasploit}\cite{metasploit} &{{Semi-automated PT process}}&{\tabincell{c}{Highly rely on the expertise}}&{\tabincell{c}{Single server}}&{\tabincell{c}{High}} \\
          \cline{2-6}
          &{Core impact}\cite{coreimpact}  & {\tabincell{c}{ Generate attack plans automatically}} &{\tabincell{c}{Inability to update dynamically}} &{\tabincell{c}{Single server \& Network topology}} &{\tabincell{c}{High}} \\
          \cline{2-6}
               &{Attack graph}\cite{graph}  &{\tabincell{c}{Attack sequence and effect are clear}} &{\tabincell{c}{Cannot interact dynamically  }} &{\tabincell{c}{Network topology}} &{\tabincell{c}{Middle}} \\
     \hline
     \end{tabular}}
\end{table}

\subsection{Penetration testing based on reinforcement learning}
In recent years, RL \cite{RL} has achieved remarkable success in game areas, e.g., Alpha Go \cite{Alphago}, Open AI Five \cite{openai5} and Alpha Star \cite{Alphastar}. Particularly, in some games the agents based on RL have surpassed human players. Just like game rules, PT is also a dynamic decision-making process based on observing the environment. The agent in RL-based PT is trained to observe and explore the dynamic network environment, so as to learn the optimal policies through trial and error \cite{trialanderror}. Intuitively, RL is an excellent option to stimulate and develop as an effective tool for penetration attacks against network security. Schwartz and Kurniawati \cite{explode}, Hu et al. \cite{autoPTDQN}, Zennaro and Erdodi \cite{modelptql} all describe PT as a POMDP, using RL to model the penetration environment and learn penetration strategies. However, the penetration agent merely observes the compromised host and the subnet, it does not observe the connection information of the global network. In other words, the agent learns the optimal strategy to execute the penetration attack through continuous trial and error learning. These studies show that RL can be applied to relatively small action spaces or CTF challenges.

Considering the observation state, the PT environment is different from the  RL game environment, because the action space of the network penetration environment is discrete and high-dimensional. It is still a long-term challenge to extend DRL to a network with a more extensive action space \cite{explode}. Arnold et al. \cite{largespace} proposed a framework called Wolpertinger, which uses the Actor-Critic (AC) framework to learn strategies in a large action space that combines sub-linear complexity. It could be unstable in a spare reward environment since the gradient cannot be back-propagated for the training of the actor network. To address the issue of sparse rewards in large action spaces, Zhou et al. \cite{zhouimDQN} proposed an improved DQN algorithm NDSPI-DQN to optimize the penetration path. It effectively reduces the agents' action space by decoupling attack vectors, but the method is only suitable for simulation scenarios. Besides, Tran K et al. \cite{HA-DRL} proposed automated PT based on deep hierarchical reinforcement learning, denoted as HA-DRL, which uses algebraic action decomposition approach to deal with the sizeable discrete action space of PT simulators. However, the space of actions will grow up exponentially with the complexity of the network. Thus the training costs are relatively high. In addition, Bland et al. \cite{bland2020}, Elderman et al. \cite{elderman2017} and He et al. \cite{he2016} proposed the application of multi-agent reinforcement learning for automated penetration testing of network security simulation scenarios, and they also proposed a learning algorithm of the optimal strategy. This method can achieve satisfying performance in simulated scenarios at the cost of high training complexity.

\subsection{Imitation learning}
Imitation learning \cite{im1} refers to automatic learning from examples provided by experts. The purpose of training the model is to fit the trajectory distribution of the strategy generated by the model with the trajectory distribution of the input. According to different ways of strategy optimization, imitation learning methods roughly divided into Behavioral Cloning (BC)  \cite{im10}\cite{im9}, Inverse Reinforcement Learning (IRL) \cite{im11} and Generative Adversarial Imitation Learning (GAIL) \cite{GAIL}.
 
BC was proposed by Torabi et la.\cite{im9} to directly learn the optimal action in the sampling state from the expert's empirical data without constructing a reward function and predict the corresponding optimal action in the new state after learning. Nevertheless, BC is prone to cascade errors. Based on BC, IRL finds a reward function that explains these strategies or behaviors by giving optimal strategies or trajectories. In order to improve the performance of IRL, Wang et al. \cite{im13} proposed Imitation Learning via Inverse Reinforcement Learning (IRL-IL) based on IRL. It is designed to solve the optimal strategy through RL, restore the expert strategy indirectly, and make a plan with long-term. Consequently, it solves the cascade error problem of BC, leading to more substantial generalization and robustness. However, the linear reward function of most IRL-IL methods contains certain restrictions \cite{im15}, i.e., the representation ability of its reward function is insufficient, and the set penalty item cannot be assigned to the expert strategy with a larger reward value as much as possible; what is more, the iterative solution of the RL sub-process requires lots of computing resources \cite{im16}. In order to make up for these limitations, Jonathan Ho et al. \cite{GAIL} combined the Generative Adversarial Network (GAN) \cite{GAN} with IRL-IL proposed GAIL. The expert data is compared with the data generated by the agent network and optimized the model reversely so that the agent can learn the strategy of approaching the expert and alleviate the training problem of IRL-IL. In addition, the algorithm has achieved a considerable performance improvement beyond the current model-free methods when simulating problematic behaviors in large-scale high-dimensional environments \cite{maxentropy}.

\section{Preliminaries\label{Preliminaries}}
This section briefly introduces reinforcement learning and its algorithm Q-learning. Besides, we also introduce DRL algorithms based on different policies, i.e., Asynchronous Advantage Actor Critic (A3C) and Distributed Proximal Policy Optimization (DPPO).

\subsection{Reinforcement learning}

RL is a machine learning method based on the sequence interaction between the agent and the environment. RL is usually modeled as a Markov process to solve sequential decision-making problems. It can be represented by a four-tuple $<S, A, P, R>$, containing the state space $S$, action space $A$, state transition probability $P$ and reward function $R$. Through continuous interaction with the environment, the agent $a_{t}$ acts under the current state $s_{t}$ according to the learned policy $\pi$. Simultaneously, the environment feeds back to the agent a scalar reward value $r\left(s_{t}, a_{t}\right)$ to evaluate the action's quality, and then transfers to the next state according to the state transition probability $P\left(s_{t+1} \mid s_{t}, a_{t}\right)$.

The objective of RL is to maximize the cumulative reward over time $t$, denoted by the reward function:
% \begin{equation}
%     \pi^{*}=\arg\max_{\pi} \sum_{t=0}^{T} \gamma^{t}E_{\pi}[r_{t}],
% \label{equ:1}
% \end{equation}
\begin{equation}
    R_{t}=\sum_{k=0}^{\infty} \gamma^{k} r_{t+k}
\label{equ:1}
\end{equation}
where $\gamma \in[0,1]$ is the discount factor used to measure the importance of current rewards to future rewards, the smaller the value, the agent only focus on the current rewards; on the contrary, the larger the value indicates that the agent will concentrate on to the future long-term returns. 

In this paper, we consider the value-based RL algorithm Q-learning \cite{DRL}, the policy-based multi-threaded DRL algorithms A3C \cite{ACA3C} and DPPO \cite{snoop33}.

 \subsubsection{Q-learning}
Q-learning is a representative algorithm in value-based model-free learning. $Q$ represents the quality function $Q(s, a)$ of the policy $\pi$, which refers to the expectation when executing action $a$ in the state of $s$ at a specific moment.
The algorithm associates all states $s$ with action $a$ to form a $Q$ table to store the $Q$ value of the state-action pair $(s, a)$, and generates the best policy $\pi ^ {*}$ by selecting the action with the highest $Q$ value in each state. The update mechanism explaining in formula:
\begin{equation}
    Q(s,a)\leftarrow Q(s, a)+\alpha[r+\gamma \max _{a^{\prime}}Q(s^{\prime},a^{\prime})-Q(s,a)]
\label{equ:3}
\end{equation}
where $s$, $a$ are the current state and action, respectively. $s^{\prime}$ is the next state that emerges as a result of action $a$, $a^{\prime}$ is a possible action in state $s^{\prime}$. $r$ is the instant reward, $\alpha$ and $\gamma$ represent learning rate and discount factor respectively.

 \subsubsection{Asynchronous advantage actor critic}
 Asynchronous Advantage Actor Critic (A3C) \cite{ACA3C} is a DRL algorithm based on policy gradient. A3C adopts multi-thread mechanism, and uses Actor-Critic(AC) network structure in both of the chief network and the thread network. AC is divided into Actor network $\pi ^ {{\theta} ^\prime} (a|s)$ and Critic network $ V ^ {{\mu} ^\prime} (s)$. The corresponding policy $\pi(a|s;{{\theta} ^\prime})$ is obtained by inputting the current state $s$, which indicates the probability of selecting an action $a$ under the condition of state $s$ and the Actor network parameters $\theta^\prime$.
 
In A3C, an advantage function $A(s,t)$ is constructed by using the output of the value function $ V(s|{{\mu} ^\prime})$ to evaluate the policy adopted, ${{\mu} ^\prime}$ is Critic network parameters. When $N$-step sampling is used, the advantage function obtained is:
\begin{equation}
% \begin{array}{l}
% \widetilde{s}=\Phi_{poison}(s), s \in [s_{1}, s_{2}, s_{3},\ldots, s_{n}]
% \end{array}
   A(s,t)=r_t+\gamma r_{t+1} + \ldots + \gamma^{n-1} R_{t+n-1}\gamma^n V(s^\prime)-V(s) = R(t)-V(s)
\label{equ:4}
\end{equation}
where $n$ means $N$-step, $\gamma$ is the discount factor. $R(t)$ represents the reward function at the current time $t$, $R(.)$ represent the value of the current state $s$. Then calculate the first derivative of $\theta ^\prime$ in Actor network and the second derivative of $\mu ^\prime$ in Critic network to update the $d\theta$ and $d\mu$, respectively.
\begin{equation}
   d\theta \leftarrow d\theta + \nabla_{{\theta}^\prime} \log
 \pi(a|s;\theta^\prime)A(s|{{\mu} ^\prime})
\label{equ:5}
\end{equation}
\begin{equation}
   d\mu \leftarrow d\mu + \partial A(s|{{\mu}^\prime})^2 / \partial {{\mu}^\prime} 
\label{equ:6}
\end{equation}
 \subsubsection{Distributed proximal policy optimization}
 
Similar to the A3C algorithm, the essence of Distributed Proximal Policy Optimization (DPPO) \cite{snoop33} algorithm uses the Proximal Policy Optimization (PPO) algorithm to train agent in a multi-thread distributed manner. The PPO algorithm is based on policy gradients and also uses a parameter $\theta$ of a neural network to approximate policy $\pi_\theta (a|s)$ directly. Besides, PPO is also implemented in an AC framework. PPO uses the advantage estimation function with reduced variance to stabilize the policy gradient while learning the approximate policy $\pi_\theta (a|s)$. The most significant difference between the AC algorithm and PPO is that three networks are implementing the PPO algorithm: a Critic-network and two Actor-networks, just the old and new Actor-networks and the primary function is to limit the update step size of the Actor-network.
  
At each training step, PPO collects experience by executing the present policy within a set of time steps, calculates the practical returns and advantages, then utilizes batch learning to optimize a clipped surrogate objective $L^{clip}$ that restrains the number of the updated strategies that may differ from the older strategies.
% \begin{equation}
%   \hat{A_t} = \delta_t + r(\gamma\lambda)\delta_{t+1}+ \ldots + (\gamma\lambda)^{T-t+1} \delta_{T-1}
% \label{equ:7}
% \end{equation}
% \begin{equation}
%   \delta_t = r_t + \gamma V(s_{t+1})-V(s_t)
% \label{equ:8}
% \end{equation}
\begin{equation}
   L^{clip}(\theta)=\mathbb{E}_t [min(r_t(\theta)\hat{A_t},clip(r_t(\theta), 1-\epsilon, 1+\epsilon) \hat{A_t} ]
\label{equ:9}
\end{equation}
where $\theta$ is the neural network parameter to approximate policy $\pi$. $\hat{A_t}$ refers to the estimation of the advantage function at the moment $t$. $\epsilon$ refers to the hyperparameter, used to restrict the change range of $r_t(\theta)$. It represents the probability ratio of the new policy to the old policy:  
\begin{equation}
   r_t(\theta)=\frac{\pi_\theta(a_t|s_t)}{\pi_{old}(a_t|s_t)}
\label{equ:10}
\end{equation}

In addition, the discrepancy between DPPO and A3C is that the update mechanisms of the chief and threads are different. The chief network of A3C does not participate in training. It only collects each thread's trained parameters, weighing them and sending them to each thread. The parameter of the child thread is new, and the parameter of the chief is old. On the contrary, the chief network of DPPO first starts training, sends its trained parameters to each thread, and then performs the process of parameter transfer and parameter update continuously, in which the threads always carry out the old strategy training and the main thread for the training of new policies.
\section{Methodology\label{Method}}
\begin{figure*}[!htp]
\centering
        \includegraphics[width=0.77\linewidth]{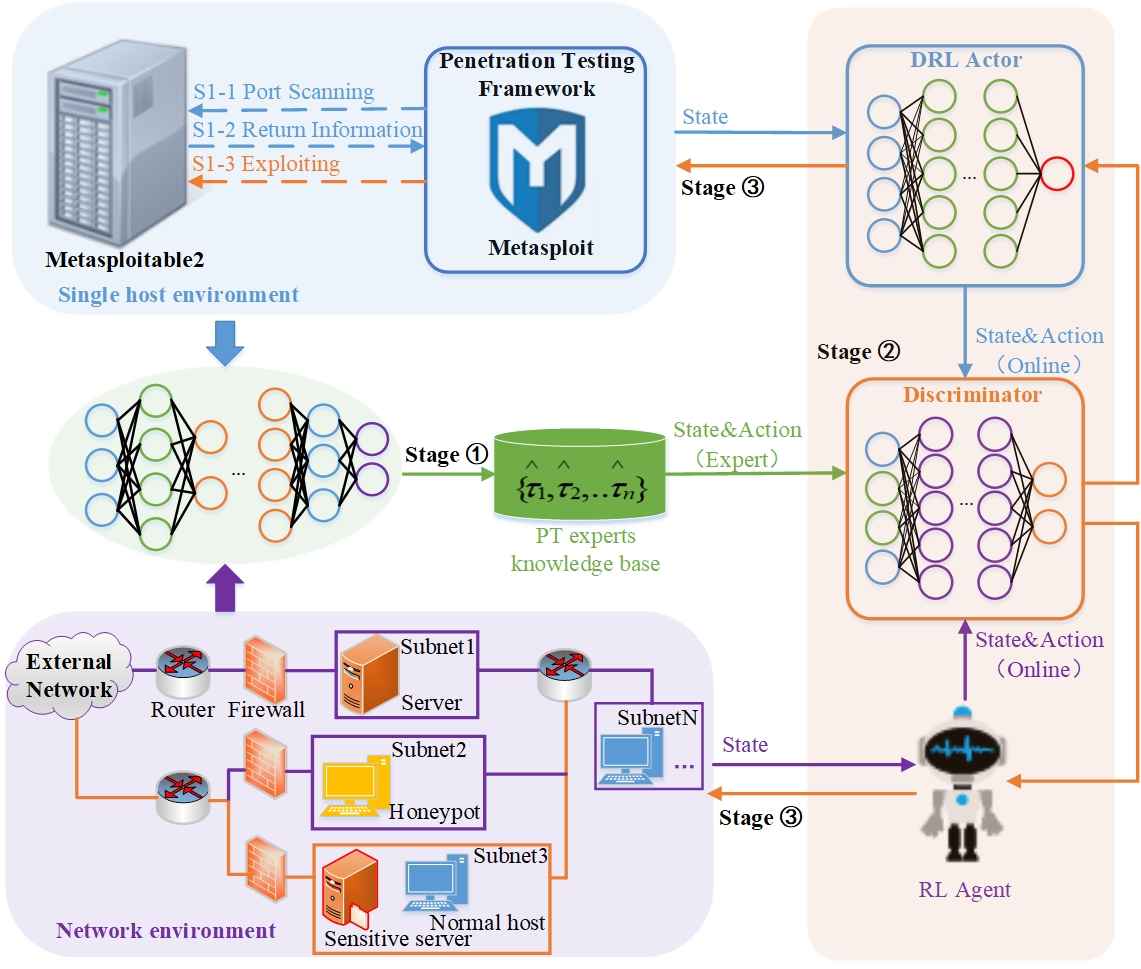}
\caption{Overall framework of GAIL-PT. Stage \textcircled{1} Construction of penetration experts knowledge base: collecting PT expert samples in different penetration scenarios, storing the state-action pairs when using RL / DRL model executing successful exploitation to construct the expert knowledge database. Stage \textcircled{2} GAIL training: Input both the expert samples and state-action pairs generated by the different RL / DRL model online into the discriminator in GAIL for training.
Stage\textcircled{3} Automate penetration testing based on GAIL: RL / DRL agent applies the new action from the well-trained discriminator to automate the PT process and execute more effective penetration attacks.}
\label{fig:1}
\end{figure*}

In this section, the method of automated penetration testing based on GAIL is introduced. First, the state-action pairs are automatically collected to construct an expert knowledge base when the pre-trained DRL / RL model executes successful post-exploit.Second, input the expert knowledge and the state-action pairs generated online by the different DRL / RL model into the discriminator of GAIL for training.At last, the discriminator's output reward is applied to guide the agent to perform the action with a higher penetration success rate to improve PT's performance. The overall framework of the proposed GAIL-PT is illustrated in Fig.~\ref{fig:1}, including three stages: \textcircled{1} Construction of penetration experts knowledge base; \textcircled{2} GAIL training; \textcircled{3} Automate penetration testing based on GAIL.

\subsection{Construction of penetration experts knowledge base}
Expert sample data are necessarily acquired in advance when conducting PT based on GAIL to train agents in different PT environments. Different from general game scenes, e.g., Gym and Atari \cite{DQN}, there are not any published expert samples for agent training in PT environment. Therefore, to our best knowledge, it is the first time, we construct penetration experts knowledge base for PT.

There are several ways to obtain PT expert knowledge samples, such as saving penetration rules defined by network security experts. It is usually difficult and expensive to collect these rules manually. Therefore, when using DRL or RL model to perform PT, we automatically collect the state-action pairs of the agent when successfully executing post-exploit, then save them into the expert knowledge base constructed online. At present, DeepExploit \cite{DeepExploit} is a better way to exploit the single target host Metasploitable2 \cite{metasploitable2} than manually using Metasploit.Using DeepExploit to perform automated PT will encounter three different results in an authentic single target host: unsuccessful exploit, successful exploit and successful post-exploit. 

\begin{figure*}[!htp]
\centering
        \includegraphics[width=0.7\linewidth]{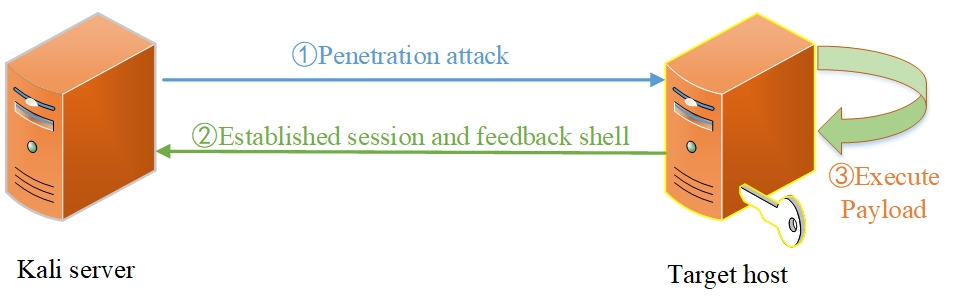}
\caption{Process of executing integral penetration attack. Stage \textcircled{1}: Kali server launches penetration attack toward Target host. Stage \textcircled{2}: Target host establishes session and feedback shell with Kail server after successful exploit. Stage \textcircled{3}: Kali server uses the payload remotely to the target Server for executing related operations to promote the privileges and obtain its root authority.}
\label{fig:2}
\end{figure*}

As shown in Fig.~\ref{fig:2}, when the Kali server initiates a penetration attack to the target host, if stage 1 executes successfully, the target server will establish a session connection with the Kali server at the same time like stage 2, then will use the payload to execute related operations on the target host to gain the root privilege. We call the process of privilege promotion post-exploit, which corresponds to stage 3. Suppose stage1, stage 2 and stage 3 all execute successfully, in that case, it indicates that the Kali server has successfully exploited the target host and performed subsequent privilege promotion operations by the payload, which means the post-exploit is also successful. We set different reward values for mixed penetration results and set the highest value for a successful post-exploit. Therefore, in the process of PT, the reward value obtained by the agent can be used to confirm the penetration result at that moment. Besides, We collected the state-action pairs with the highest reward value as the expert sample data and stored them into expert knowledge base in the form of one-to-one correspondence between states and actions.

Like the single target host scenario, the mode of collecting expert sample data in the network scenario also means storing the state-action pairs with the highest reward values when exploiting the target sensitive hosts successfully at the lowest cost. 
Generally speaking, construction of penetration experts knowledge base is similar for both exploiting a single target host and exploiting a network. Specifically, exploiting a host means acting some related operations on the target host, like executing the payload, etc., finally gaining control of the target host. In the network scenario, after gaining the control of one host, the attacker can use the compromised host as a stepping stone according to network connection relationships to carry out lateral movements for obtaining control of the final sensitive host. In other words, exploiting a network is just like finding an optimal attack path, and the path shows the exploit relationship between some hosts in different subnets. It is worth mentioning that the most significant distinction between our approach and the attack graph lies in our method does not need to know the network topology information in advance but finds an optimal attack path through continuous trial and error training. 

\begin{figure*}[!htp]
\centering
        \includegraphics[width=0.68\linewidth]{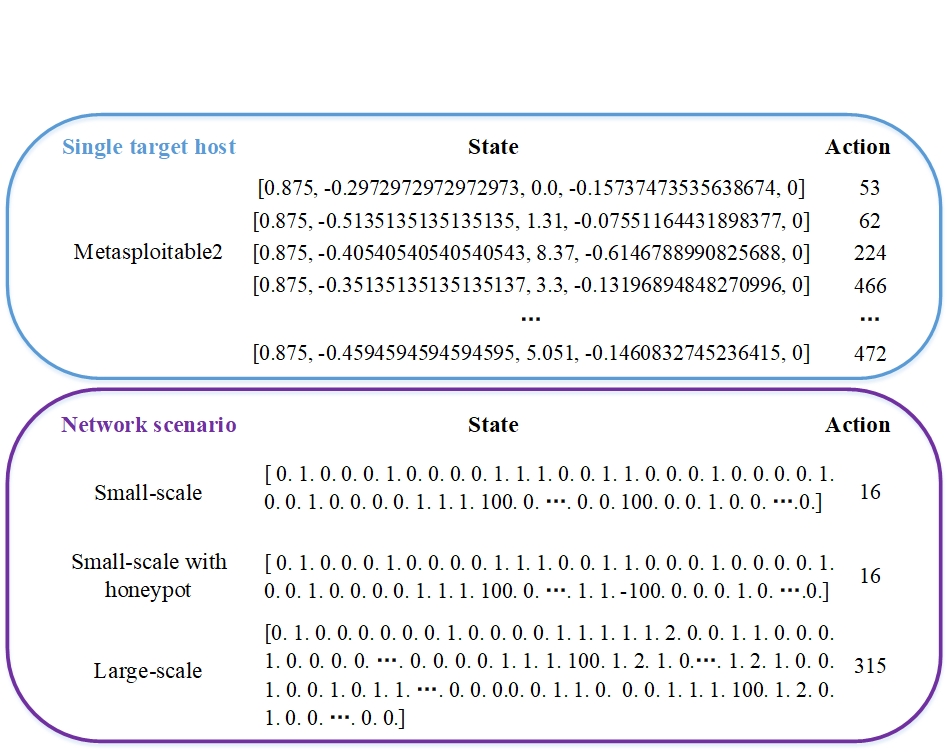}
 \caption{Examples of PT expert knowledge base.} 
\label{fig:3}
\end{figure*}

For better understanding, an illustration example is shown in Fig.~\ref{fig:3}. There are various combinations of state-action pairs due to the diversity of vulnerabilities when exploiting the real target host Metasploitable2. The five dimension state represents operating system, port service, product version, protocol and exploit module type. The action corresponds to the number of payloads in Metasploit. Take the first row of data of a single target host as an example; 0.875 means the operating system is Linux, the port service is SSH (Secure Shell),  the product version is 0.0, the protocol is TCP (Transmission Control Protocol), and the exploit module type is 0 represents choosing the first item of the exploit list. As we know, there are 593 (0 $\sim$ 592) payloads in Metasploit. Action 53 means choosing the 53rd payload to execute the exploit operation.

However, for expert sample data in the network scenario, its state represents the configuration details and vulnerability information of all hosts observed by the agent in the different network environments. Besides, action means the operations of vulnerability scanning, exploit and privilege promotion executed by the agent arriving at the target host. Take the small-scale scenario's state and action as an example. The state vector consists of different host vectors for each subnet and starts with the first host. At the beginning of the state vector, the host location is represented by a 10 bit one-hot encoding. The numbers, i.e., 0 means false and l means true, behind successively represent compromised, reachable, discovered, value (specific value of different hosts), discovery value, access, os (operating system, i.e., 1 means Windows, otherwise means Linux), exploit services and privilege promotion processes. Besides, the additional final row in the state vector, representing action success, connection error, permission error and undefined error. It also uses 1 or 0 to represent true or false. 

There are eight hosts in the small-scale network scenario (the scenario is shown in Fig.\ref{fig:smallscale}), including two types of the operating system: Windows and Linux, three types of exploit services: SSH, HTTP (Hyper Text Transfer Protocol), and FTP ( File Transfer Protocol ), two types of privilege promotion process: Tomcat and daclsvc, two types of firewall restrictiveness: HTTP and SSH. Consequently, the actions could have 72 (8$\times$9) types of operation combinations. The 16th action means executing privilege promotion by the Tomcat process on the sensitive hosts in subnet 2, which is critical in exploiting the small network scenario. In short, the state-action pair corresponding to the best penetration attack path, and is unique and deterministic, so the expert sample data is relatively fixed. After that, we only did simple and repeated expansions of the sample size.

\subsection{GAIL training}
After collecting the PT expert sample data, this section introduces training the GAIL network. 

Our proposed GAIL contains three different neural networks: Policy Neural Network (Actor), Value Neural Network (Critic) and Discriminator Network (Discriminator). The Actor consists of a four-layer network. The hidden layers comprise three dense layers with output dimensions of 50, 100 and 200 connected through the ReLU activation function. The last is the output layer with a Softmax activation function, and the output dimension corresponds to the number of actions. 

The network structure of Critic is similar to that of Actor. The only difference is that the last output layer is linear, the dimension is one. Discriminator, known as Reward Neural Network, has a four-layer network with three hidden layers of the same structure as the Actor's hidden layers. The last output layer is activated by Sigmoid and outputs a specific probability value as a discounted reward.

\begin{figure*}[!htp]
\centering
        \includegraphics[width=0.72\linewidth]{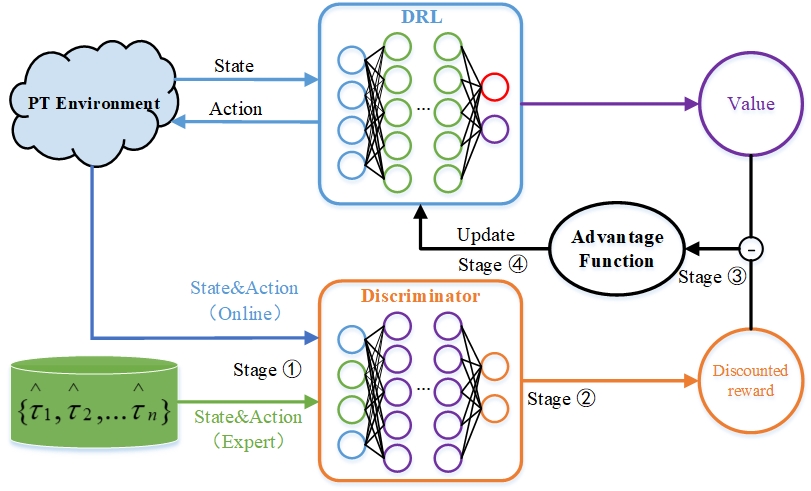}
\caption{The Process of GAIL training. Stage \textcircled{1}: Put the state-action pairs from PT expert knowledge base and generating by online model respectively into the Discriminator for training together. Stage \textcircled{2}: Through maximizing the standard action reward given by the sample of PT experts and minimizing the action reward output by the agent to train the Discriminator. Stage \textcircled{3}: Subtract the discount reward and the value output by the Discriminator and the different DRL model to get the advantage function. Stage \textcircled{4}: Use advantage function to update DRL model meanwhile guiding agent output actions with a higher probability of successful penetration.}
\label{fig:4}
\end{figure*}

The process of GAIL-training is exhibited in Fig.~\ref{fig:4}. The Actor, Critic and Discriminator are alternately trained. The inputs of the Actor and Critic are both currently observed states, while the output of Actor is the corresponding action, and the output of Critic is the value of the current state. The input of the Discriminator are the state-action pairs of the agent and PT experts. Besides, the state-action pairs of the expert sample are only used for training, and the output of the discriminator is the discount reward value.

The main processes of GAIL-training are as follows: (1) Put the state-action pairs from PT expert knowledge base and generating by DRL / RL online model respectively into the Discriminator for training together. (2) Through maximizing the standard action reward given by the sample of PT experts and minimizing the action reward output by the agent to train discriminator. (3) Use the discriminator's output to replace the original model's reward function to guide the training directions. (4) The DRL model output predicted distribution is infinitely close to the trajectory of PT expert knowledge base as much as possible.

During the training process of imitating PT experts strategy, the Actor network is used to replace the generator $G$, its output action and state are paired into the Discriminator, and compared with the expert data. The output of the Discriminator $D:S \times A\rightarrow(0,1)$ is considered as the reward signal to guide strategy learning in imitation learning. Therefore, regarding the training of policy and reward function as the process of a game that is analogous to the game of $G$ and $D$ in Generative Adversarial Network (GAN), the objective function $L_{GAIL}(\pi,D)$ as follows:
\begin{equation}
   \min_{\pi}\max_D L_{GAIL}(\mathbb{E}_{\pi}[\log D(s,a)]+\mathbb{E}_{\pi_E}[\log(1-D(s,a))]
\label{equ:11}
\end{equation}
The $\pi$ in the formula represents the policy that should be learned, and $\pi_{E}$  means the expert strategy. The first term $\log D(s, a)$ denotes the judgment of the Discriminator on the factual data, and the second term $\log(1-D(s, a))$ denotes the judgment of the generated data. Through such a maximum-minimum game process, $G$ and $D$ are optimized alternately to train the Actor and Discriminator networks. In the training process, we get the total loss by adding the expert loss and the agent loss, minimizing the loss function by gradient derivation to update the Discriminator and Actor network parameters in reverse:
\begin{equation}
   Loss =\min \sup_{D\in(0,1)^{S\times A}} \mathbb{E}_{\pi}[\nabla_{\omega}\log D_{\omega}(s,a)]+\mathbb{E}_{\pi_E}[\nabla_{\omega}\log(1-D_{\omega}(s,a))]-\lambda\nabla_{\theta}H(\pi_{\theta})
\label{equ:12}
\end{equation}
where $H(\pi_{\theta})\overset{\Delta}= \mathbb{E}_{\pi}[-\log \pi_\theta(a|s)]$, it represents the entropy of the imitation policy, controlled by a constant $\lambda(\lambda \ge 0)$, as the policy regular term in the loss function, and $\omega$ and $\theta$ signifies the parameter of Discriminator and Actor network separately.
\subsection{Automated penetration testing based on GAIL}
After GAIL's training is completed, automated penetration testing can be performed based on GAIL. This section narrates the automated penetration testing process based on GAIL. The training framework has been shown in Fig.~\ref{fig:1}. 
\subsubsection{Automated penetration testing based on GAIL combined with DRL}
In the actual scenario of exploiting the single target  Metasploitable2, we define the PT process as an MDP at first. The states $s$ of the agent is defined by the information obtained by Nmap through port scanning; secondly, we define the actions $a$ selected by the agent in the PT, the actions corresponding to the payload sequences list in Metasploit Framework (MSF); at last, the reward $r$ is set according to the penetration result of the applied payload. We use four different DRL models to train the agent, such as A3C, DPPO, A3C-GAIL and DPPO-GAIL. What is more, we all used a multi-threaded mechanism for training to speed up the learning efficiency of the agent.

Each of the A3C and DPPO models we used consisted of 4 layers neural networks: the input layer, the hidden layers with three fully connected and the output layers. The state inputs of the model, i.e., operating system, port service, product version, protocol and exploit module type, are all obtained by scanning the target through the vulnerability scanning tool Nmap. The action output of the model corresponds to the probability distribution and the state value of each payload in Metasploit. For the Actor network in the model, we use Softmax as the activation function, and the output of the Actor network is the probability distribution $P(a)$ of each action; while the Critic network uses Linear as the activation function, and its output is the state value $V(s)$ corresponding to the current state.

We introduced a discriminator to participate in the Actor and the Critic network training during the Actor and the Critic network training. There are specific differences from A3C and DPPO models in the training process based on the DRL model combined with GAIL. The difference between the value function and the discriminator output discount reward is applied as the advantage function to guide the training of the policy network, and the output of the GAIL model is consistent with the original model. 

The DRL model is regarded as an attacker in the learning process. According to the input vulnerability information, the agent will choose an effective payload to  exploit the target host until further getting its root authority.
Therefore, obtaining the root authority of the target host is a sign of a successful exploit. To be more precise, it is the post-exploit after the privilege promotion operation has been done, and then the agent will be rewarded. In addition, the value of rewards $r$ depends on whether the PT attack is effective. 

As shown in Fig.~\ref{fig:2}, if the agent executes a penetration attack from the Kali server to the target host, establishing a session connection with the Kali server and then executing the payload to gain root authority of the target host favorably. It indicates that the post-exploit is prospering, and the agent will obtain the highest reward.
On the contrary, if the agent only initiates the penetration attack without establishing a connection with the Kali server, the agent will be given a lower reward at this time. Furthermore, if the agent has not exploited the target host successfully, its reward will be negative. The specific settings of the reward are as follows:
\begin{itemize}
\item The post-exploit is successful, $r$=$100$;
\item The exploit is successful, but the root authority for target host is not obtained, $r$=$1$;
\item The exploit is failure, $r$=$-1$;
\item For other operations, regardless of whether the exploitation is successful or not, $r$=$-1$.
\end{itemize}

Negative reward $r$=$-1$ means to punish the agent, and the agent tries to maximize the reward all the time. Therefore, if the agent has been punished with negative rewards, it will try to reach the penetration goal as soon as possible. It corresponds to a situation where an actual attacker realizes detecting vulnerabilities and exploiting them successfully. In addition, as long as the agent gets a reward, it symbolizes that each round of training is over.

Details regarding the automated PT via GAIL combined with DPPO are presented in \textbf{Algorithm 1}.
\begin{algorithm}
	\caption{DPPO-GAIL}
	\begin{algorithmic}[1] %
		\Require Expert PT trajectories $\hat{\tau}_{GAIL}\sim{\pi_{E}}$, five-dimension state $s$ including operating system type, opening-port service name, product version, penetration module type, and key information of exploiting target type from using Namp to scan Metaspolitable2, iterations $i=200000$.
          \Ensure Action $a$ means payload which is infinitely close to the expert samples $a_E$.
          \State Initialize DPPO actor-net and GAIL discriminator-net parameters $\theta_0$; $i_0$.
         \For{$i = 0,1,2,\ldots$}
          \State Generate state-action pairs from DPPO model online and store them as $\pi_t$. 
         \State  Sample online trajectories $\hat{\tau}_{i}\sim{\pi_{\theta_i}}$.
         \State  Update the discriminator-net parameters from $\omega_{i}$ to $\omega_{i+1}$ with the gradient:
         \State  $G_d = \hat{\mathbb{E}}_{\pi_{GAIL}}[\nabla_{\omega}\log D_{\omega}(s,a)]+\hat{\mathbb{E}}_{\pi_t}[\nabla_{\omega}\log(1-D_{\omega}(s,a))]$. 
         \State From DPPO actor-net take a policy step from $\theta_i$ to $\theta_{i+1}$ respectively.
         \State Using the TRPO rule with the cost function $\log(D_{\omega_i+1}(s,a))$.
         \State Specifically, DPPO takes a $L^{clip}$ natural gradient step to update old and new policy with \eqref{equ:9}, and: 
         \State  $\hat{\mathbb{E}}_{\pi_{GAIL}}[\nabla_{\theta}\log \pi_{\theta}(a|s)Q(s,a)-\lambda\nabla_{\theta}H(\pi_{\theta})]$,
         \State  where $Q(\bar{s}, \bar{a})=\hat{\mathrm{E}}_{\pi_{GAIL}}\left[\log \left(D_{\omega_{i+1}}(s, a)\right) \mid s_{0}=\bar{s}, a_{0}=\bar{a}\right]$
         \EndFor
	\end{algorithmic}
\end{algorithm}

\subsubsection{Automated penetration testing based on GAIL combined with RL}
When conducting penetration testing in the network attack simulator NASim\cite{nasim}, we also model the PT process as an MDP at first, the state is defined as the configuration details and vulnerability information of all hosts observed by the agent in the different network environments. The larger the number of hosts, the larger the state scale. Besides, action means the manipulations of vulnerability scanning, exploit and privilege promotion executed by the agent arriving at the target host. We defined the action as an attack vector, $<m,c>$ means the agent manipulates $m$ on the host computer $c$. In each training step, the scale of operating space for an agent reaches $(P\times Q)$, where $P$ are the number of hosts in the network, $Q$ represents the number of manipulations that exploit the final target host could perform. After the agent executes a particular operation, the environment will feedback a corresponding reward value. The reward value is calculated as shown in \eqref{equ:13}, including the value of the host and the cost of exploits, privilege promotion, or scanning operations on the host.
\begin{equation}
R=\sum_{c\in C} value(c) - \sum_{m\in M} cost(m)
\label{equ:13}
\end{equation}

We use the RL algorithm Q-learning as a benchmark to automatically generate penetration paths and optimize its process for path-finding. During the training process, the agent will select the corresponding representative vulnerabilities according to the service to calculate the vulnerabilities' probability. $C$ in Equation \eqref{equ:14} represents the set of host computers in different networks that the agent can exploit, and $M$ represents the set of manipulations the agent can adopt. The attacker intends to maximize the cumulative reward and explore the optimal penetration path, until exploiting the most valuable sensitive target hosts successfully with as few operations as possible.
\begin{equation}
\max_\pi{\mathbb{E}}[\sum_{t=0}^{\infty}\gamma^t R(s_t,a_t,s_t+1)|\pi]
\label{equ:14}
\end{equation}

Subsequently, we combined GAIL with the Q-learning algorithm. The Q-learning-GAIL algorithm is applied to three different scenarios of a small-scale network (considering whether there exists a honeypot) and a large-scale network to train the agent to accomplish the optimization of finding the penetration path. The Q-learning algorithm is different from the DRL algorithm. Its optimization process is to continuously minimize the difference $TD-error$ between the target value $Q_{tar}$ and the realistic evaluation value $Q_{eva}$. When calculating the current value, $Q_{eva}$, the immediate reward $r$ needs to be are summed with the discounted maximum target value $Q_{tarnex}$ in the next state. Therefore, to improve the optimization effect, we superimpose the instant reward $r$ and the reward $r_d$ of the discriminator output, which is used to guide the learning of the imitation policy, so that leads the action distribution of the RL model is close to PT experts samples as much as possible.

Details regarding the automated PT with GAIL combined with Q-learning are presented in \textbf{Algorithm 2}.
\begin{algorithm}
	\caption{Q-learning-GAIL }
	\begin{algorithmic}[1] %
		\Require Expert PT trajectories $\hat{\tau}_{GAIL}\sim{\pi_{E}}$, states $s$ including configuration knowledge and vulnerability information of all hosts, iterations $i=20000$.
		\Ensure Actions $a$ means exploits or privilege escalation which are infinitely close to the expert samples $a_E$.
		\State Initialize $Q(s,a)$ arbitrarily and GAIL discriminator-net parameters $\omega_0$.
		\For{$i = 0,1,2,\ldots$}
		\State Generate state-action pairs from RL model online and store them as $\pi_t$. 
	 \State Sample online trajectories $\hat{\tau}_{i}\sim{\pi_{\theta_i}}$, initial $s$..
		\State Update the discriminator-net parameters from $\omega_{i}$ to $\omega_{i+1}$ with the gradient:
	\State $G_d$ = $\hat{\mathbb{E}}_{\pi_{GAIL}}[\nabla_{\omega}\log D_{\omega}(s,a)]+\hat{\mathbb{E}}_{\pi_t}[\nabla_{\omega}\log(1-D_{\omega}(s,a))]$. 
		\State Choose $a$ from $s$ using derived from $Q($$\epsilon$-greedy$)$.
		\State Using the TRPO rule with the cost function $\log(D_{\omega_i+1}(s,a))$.
		\State Take action $a$, observe $r$, $\overline{s}$. 
		\State From discriminator-net get discounted reward $r_d$.  
		\State Update $Q(s,a)$$\leftarrow Q(s,a) + \alpha[r+r_d+\gamma\max_{\overline{a}}$ Q($\overline{s}$,$\overline{a}$)$-Q(s,a)], $s$\leftarrow$ $\overline{s}$. 
		\EndFor
	\end{algorithmic}
\end{algorithm}

\subsection{Convergence analysis of GAIL}
Before GAIL \cite{GAIL} was proposed, Ho et al. \cite{im15} first proposed the inverse reinforcement imitation learning (IRL-IL) framework:
\begin{equation}
 \max_{\pi}\min_r\mathbb{E}_{\pi}[r(s,a)]-\mathbb{E}_{\pi_E}[r(s,a)]+\psi(r)
 \label{equ:16}
\end{equation}
where $r(s, a)$ represents the reward function, and the reward $r$ corresponding to the state-action pair $(s, a)$, ${\pi_E}$ represents expert strategy, $\pi$ represents the strategy need to be learned, $\mathbb{E}_{\pi}[$·$]$ denotes the expectations about the strategy, and $\psi(r)$ is the penalty term of the reward function. Based on this framework, GAIL gave a particular but more reasonable form of penalty term:
\begin{equation}
\psi_{G A I L}(r) \triangleq\left\{\begin{array}{cc}
\mathbb{E}_{\pi_{E}}[g(r(s, a))] & \text { if } r>0 \\
+\infty & \text { otherwise }
\end{array}~g(x)=\left\{\begin{array}{cc}
x+\log \left(1-e^{-x}\right) & \text { if } x>0 \\
+\infty & \text { otherwise }
\end{array}\right.\right.
\end{equation}
% \begin{equation}
% \begin{aligned}
% \psi_{G A I L}(r) &\triangleq\left\{\begin{array}{cc}
% \mathbf{E}_{\pi_{E}}[g(r(s, a))] & \text { if } r>0 \\
% +\infty & \text { otherwise }
% \end{array}\right.\\
% g(x)&=\left\{\begin{array}{cc}
% x+\log \left(1-e^{-x}\right) & \text { if } x>0 \\
% +\infty & \text { otherwise }
% \end{array}\right.
% \end{aligned}
% \end{equation}
The valuable feature for the penalty term $\psi_{GAIL}(r)$ is that it encourages the new reward function $g(x)$ to assign a more considerable reward value to the expert strategy ${\pi_E}$ ; its special feature is that if the reward function satisfies a specific form:
\begin{equation}
r(s,a)=-\log D(s,a)
\label{equ:17}
\end{equation}
where $D(s,a)$ represents the probability that the Discriminator $D$ discriminates $(s, a)$ produced by the expert strategy. It happens to combine IRL-IL with Generative Adversarial Networks (GANs): the Actor network of inputting state and generating policy can be compared to Generator $G$. The reward function of inputting state-action pairs and outputting discount rewards can be compared to Discriminator $D$. The optimization process of the policy based on the current reward function can be analogous to the training process of the Generator. Similarly, the optimization process of the reward function can be analogous to the training process of the Discriminator.

Therefore, the core of GAIL is to minimize the Jensen-Shannon divergence between the state-action sample distribution $\rho_{\pi}(s, a)$ generated by the strategy $\pi$ and the expert sample distribution $\rho_{\pi_E}(s, a)$:
\begin{equation}
\psi^*_{GAIL}=\sup_{D\in(0,1)^{S\times A}} \mathbb{E}_{\pi}[\log D(s,a)]+\mathbb{E}_{\pi_E}[log(1-D(s,a))].
\label{equ:18}
\end{equation}

Equation \eqref{equ:18} proportional to the best negative logarithm loss of the binary classification problem of distinguishing the different state-action pairs. The results show that the optimal loss is the Jensen-Shannon divergence under a constant offset and scale, which is also the square measure between the standardized occupancy distributions:
\begin{equation}
D_{JS}(\overline{\rho}_{\pi},\overline{\rho}_{\pi_E})\overset{\Delta} = D_{KL}(\overline{\rho}_{\pi}||(\overline{\rho}_{\pi}+\overline{\rho}_E)/2)+D_{KL}(\overline{\rho}_{E}||(\overline{\rho}_{\pi}+\overline{\rho}_E)/2)
\label{equ:19}
\end{equation}
where $\overline{\rho}_{\pi}=(1-\gamma)\rho_{\pi}$, $\overline{\rho}_{\pi_E}= (1-\gamma)\rho_{\pi_E}$. Regarding the causal entropy as a policy regular item, it is controlled by $\gamma (\gamma\geq 0)$, and in order to make the classification clearer, abandoning the standardization of the occupancy measurement $1-\gamma$, thereby obtaining a new imitation learning algorithm:
\begin{equation}
\mathop {\min }_{\pi}\psi^*_{GAIL}({\rho}_\pi-\rho_{\pi_E})-\lambda H(\pi)=D_{JS}({\rho}_\pi,\rho_{\pi_E})-\lambda H(\pi)
\label{equ:20}
\end{equation}

So we can get a strategy, and its occupancy rate indicator minimizes the Jensen-Shannon divergence between the generated and expert samples. The above Equation \eqref{equ:20} minimizes the accurate metric between the occupancy metrics. Therefore, unlike the linear apprentice learning algorithm, it can imitate expert strategies accurately. It also proves that the state-action pair generated by the instant strategy $\pi$ can fit the sample distribution of the expert strategy $\pi_E$ finally by minimizing the difference between ${\rho}_{\pi}(s, a)$ and ${\rho}_{\pi_E}(s, a)$ so that GAIL could come to converge.

\subsection{Analysis of algorithm complexity}
In the realistic scenario of exploiting Metasploitable2, the automated PT process is completed based on A3C-GAIL and DPPO-GAIL. As for DRL algorithms, the algorithm's complexity is related to the number of training episodes $m$ and maximum steps $t$ per episode. For the A3C algorithm, the thread network and the chief network adopt AC structure with only one for-loop statement nested. Besides, the GAIL network also nests one for-loop. Because of the GAIL, the Actor and the Critic adopt an alternative train mechanism. After combining with the GAIL, the algorithm complexity of A3C-GAIL is $O(n_mn_t)$. In addition, the DPPO algorithm uses the new and old policy update mechanisms. The training program of the chief network contains three for-loop statements, two of which are nested in another for-loop in a juxtaposed relationship. One of these two for-loop statements are used for model training. The other is used for agent choosing action. Then combined with GAIL, the algorithm time complexity of DPPO-GAIL is $O(n_mn_t)$.

The Q-learning algorithm's complexity is relatively small for the network scene to automatically find and optimize penetration paths because there is only one for-loop statement in it. Consequently, after nesting with GAIL, the algorithm time complexity of Q-learning-GAIL stays at $O(n)$.

\section{Experiments\label{Exp}}
The experiments we conducted are divided into the following two parts. First, we test the penetration performance of using the combination of GAIL network and DRL algorithms A3C-GAIL, and DPPO-GAIL to automate PT in the real Metasploitable2 scenario. Then, considering the complexity of the actual penetration environment, we employ the network attack simulator NASim to find the optimal penetration path automatically. To testify the effectiveness of the RL model, we combine GAIL network with RL algorithm Q-learning, and evaluate the penetration performance of using the Q-learning-GAIL in a small-scale network (considering whether there exists honeypot \cite{honeypot1} \cite{honeypot2}) and a large-scale network.

Specifically, all experiments will perform on a single server with an Intel i7-7700K CPU running at 4.20GHz, 64 GB DDR4 memory, 4 TB HDD and 2 TITAN Xp 12 GB GPU card, operation system is Ubuntu 16.04, and employ Python 3.6 with tensorflow-2.1.0 and Python 3.7 with torch-1.5.0 for exploiting in metasploitable2 and NASim. 

\subsection{Research questions}
The experiment will be analyzed based on different experimental scenarios around the following four research questions (RQs):

\textbf{RQ1:} Can expanding expert knowledge samples improve penetration performance in exploiting Metasploitable2? What is the effect of the different amounts?

\textbf{RQ2:} How efficient is GAIL's penetration when exploiting Metasploitable2? Can GAIL achieve the SOTA performance?

\textbf{RQ3:} Does expanding the expert sample amount still improve the penetration performance in different network scenarios?

\textbf{RQ4:} Can GAIL also achieve the SOTA penetration performance in different network scenarios?

\subsection{Single host senario}
\subsubsection{Explanation for Metasploitable2 }
We used Metasploitable2 as the authentic single target host. The Metasploitable2 equipped an elaborated Ubuntu operating system designed as a security tool to test and demonstrate the environment for common vulnerabilities and attacks. Its primary purpose is to use as a target for launching penetration attacks by MSF. The target host opened 23 ports, and some of them opened port vulnerability information is shown in Table~\ref{tabel:metasploitable2}, including some high-risk ports such as 21, 23, 8180, etc. Many unpatched high-risk vulnerabilities include Samba, MSRPC, Shell and Command injection vulnerabilities, etc. Besides, some vulnerabilities open various services to the outside environment, and the database allows external connections. The user passwords in the system are vulnerable, and it equips with web vulnerability drill platforms such as DVWA and Mutillidae.

 \begin{table}[!htp]
\footnotesize
\centering
\caption{Part of the opened port information of Metasploitable2}
\label{tabel:metasploitable2}
\resizebox{0.3\linewidth}{!}{
\begin{tabular}{lll}
\toprule \hline
\textbf{Port} & \textbf{State} &\textbf{Service} \\ \hline
21/tcp            & open              & FTP   \\
22/tcp            & open             & SSH    \\
23/tcp            & open             & Telnet \\
\dots             & \dots            &\dots \\
8180/tcp            & open           & HTTP \\
\hline \bottomrule
\end{tabular}
}
\end{table}
\subsubsection{Experiment setting}
In order to evaluate the penetration performance of different algorithms in exploiting the target Metasploitable2, we first choose DeepExploit as the benchmark, which automates PT based on the A3C algorithm.Second, DPPO is different from A3C only in the update mechanism of the thread and chief network, so we also explored the feasibility of using the distributed algorithm DPPO to automate the PT process, and tested its performance on exploiting the same target.At last, we combined the GAIL network with A3C and DPPO algorithm, respectively. We Applied the A3C-GAIL and DPPO-GAIL algorithms to automate the PT process, exploited the target host Metasploitbale2, and tested its penetration performance.

This section uses the above four different algorithm models to automate PT, and the ultimate goal is to execute the post-exploit successfully. The four algorithms all use 20 threads for 200,000 episodes of training and the same hyper-parameter value settings, details are shown in Table~\ref{tabel:hyper-parameter}.
 \begin{table}[!htp]
\footnotesize
\centering
\caption{Hyper-parameter values of four algorithms.}
\label{tabel:hyper-parameter}
\resizebox{0.4\linewidth}{!}{
\begin{tabular}{ll}
\toprule \hline
\textbf{Hyper-parameter}  & \textbf{Value} \\ \hline
Max steps per episode              & 30  \\
Max update step size per episode   & 5 \\
Learning rate                       & 0.0001 \\
Greedy rate                        &0.8 \\
Entropy coefficient               &0.01    \\
Loss coefficient of Critic-net    &0.05    \\
\hline \bottomrule
\end{tabular}
}
\end{table}
\subsubsection{Evaluation metrics}
In the single target host scenario, we apply the following five indicators from the perspectives of DRL performance and penetration performance to analyze the answers of \emph{\textbf{RQ1}} and \emph{\textbf{RQ2}}:
\begin{itemize}
\item \textbf{Total reward.} For the training of DRL models, the reward value most intuitively reflects the learning quality of the agent, so we apply the total reward as the first indicator. The total reward is summed in every episode, which can be used to evaluate the overall performance of the algorithm;

\item \textbf{Loss value.} For the training of DRL models, we can judge the stability of the model training by the loss value, so we apply the loss value as the second indicator. The loss value is calculated when the chief and child network parameters are updated, which can also be used to evaluate the overall performance of the algorithm;

\item \textbf{Post-exploit count.} In the process of exploiting the Metasploitable2, post-exploit means the agent exploited the vulnerabilities service in a different open port and successfully executed the payload to gain the privilege for the target host. The more successful post-exploit, the more effective the PT is. It is the most intuitive reflection of penetration performance;

\item \textbf{Time cost.} In the process of exploiting the Metasploitable2, time cost means the time spend in 200000 episodes training for using DRL models to exploit the target host. It reflects the cost we should pay. The less the time cost, the better the penetration performance is;

\item \textbf{Time cost under exploiting limited vulnerabilities.} Considering the limitations of applying the MSF to exploit the target host Metasploitable2, no matter using the MSF manually or automatically to exploit Metasploitable2, we could only exploit the vulnerabilities of 11 types of services: VNC (Virtual Network Computing), Telnet, SSH, RPC (Remote Procedure Call), ProFTPd, PostgreSQL (Postgre Structured Query Language), PHP (Hypertext Preprocessor), MySQL (My Postgre Structured Query Language), IRC (Internet Relay Chat), HTTP (Hypertext Transfer Protocol), Apache. Therefore, we compared the time to exploit these vulnerabilities and used it as the last indicator to evaluate the penetration performance.
\end{itemize}

\subsubsection{Research question 1}

\begin{center}
\fcolorbox{black}{gray!20}{\parbox{0.97\linewidth}
    {
        Can expanding expert knowledge samples improve penetration performance in exploiting Metasploitable2? What is the effect of the different amounts?
    }
}
\end{center}

We will answer the question through analyzing the experiment results from the two indicators: Post-exploit count and Time cost.

This section used DeepExploit as the benchmark and then built an expert knowledge base by collecting expert state-action pairs when the post-exploit is successful. We tested the count of post-exploit and time cost of applying A3C-GAIL performing automate PT with 10,000, 25,000, 40,000, 55,000 and 70,000 pairs of five different expert sample amounts in the same training episodes, using the two evaluation indicators to overview the impact of the expanding expert samples on the penetration performance.
\begin{figure*}[!htp]
\centering
        \includegraphics[width=0.66\linewidth]{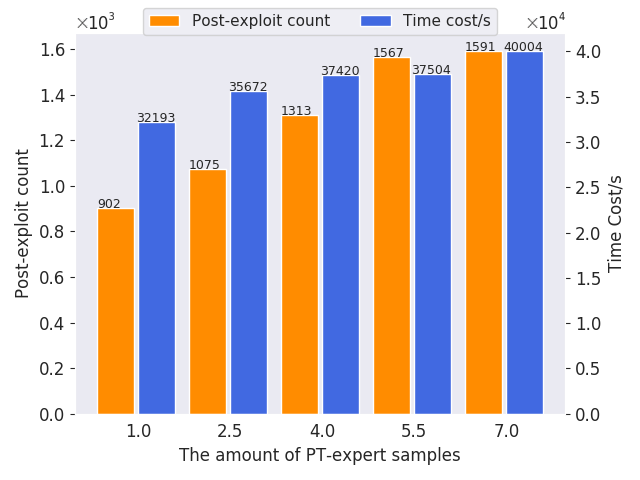}
\caption{The impact of different expert sample sizes on penetration performance in exploiting the target host Metasploitable2 }
\label{fig:amount}
\end{figure*}

It can be observed from the experiment results in Fig.~\ref{fig:amount} that under the premise of considering the count of successful post-exploit and time cost comprehensively, the increase in the sample size of experts from 10,000 to 55,000 pairs, the improvement in penetration performance positively correlating with the size of the expert samples. When the expert sample size is expanded to 70,000 pairs, although the count of successful post-exploit increases slightly, the time cost is higher. Therefore, we determined that the sample size is around 55,000 pairs to achieve better penetration performance. Besides, we also use the DPPO-GAIL model with this expert sample size to automate PT in the subsequent experiment,  and the final expert sample size is 57296 pairs.

\vspace{-0.3cm}
%%%%%%%%%%%%%%%%%%%%%%%%%%%%%%%%%%%%%%%% Answer1
\begin{center}
\fcolorbox{black}{white!20}{\parbox{0.97\linewidth}
    {
        \emph{\textbf{Answer to RQ1}}:
        Penetration performance is improved by expanding the amount of expert samples in the range of 55,000 pairs, but it tends to be saturated when the expert samples exceed this range.
    }
}
\end{center}

\subsubsection{Research question 2}
\begin{center}
\fcolorbox{black}{gray!20}{\parbox{0.97\linewidth}
    {
        How efficient is GAIL's penetration when exploiting Metasploitable2? Can GAIL achieve the SOTA performance?
    }
}
\end{center}

We will answer the question through analyzing the experiment results from the above proposed five indicators: 
Total rewards, Loss value, Post-exploit count, Time cost, Time cost under exploiting limited vulnerabilities.

\begin{figure}[t]
\centering
\subfigure[Comparison of reward]{
        \includegraphics[width=0.48\linewidth]{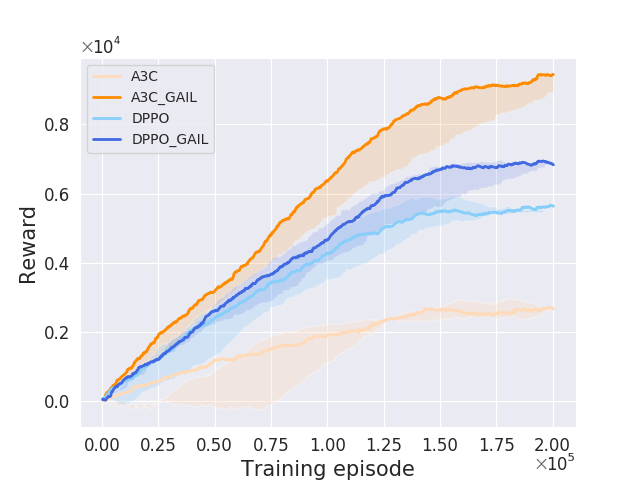}
    }
\subfigure[Comparison of loss value]{
    \includegraphics[width=0.48\linewidth]{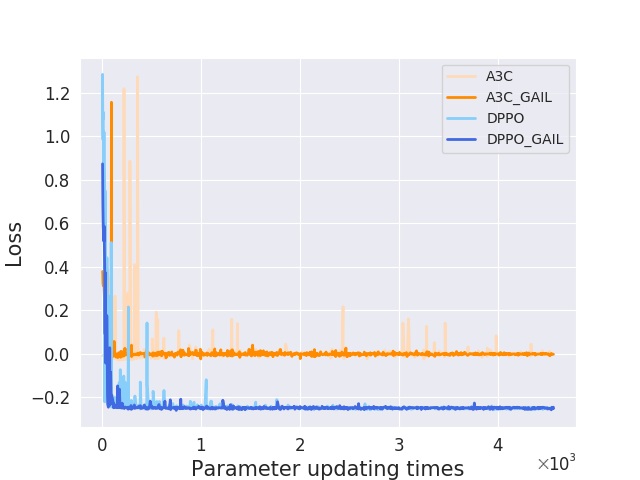}
    }
    \vspace{-0.4cm}
  \caption{Different algorithms compared the reward value (left) and loss value (right). The training episodes of the four algorithms are all 200,000, they are all trained five times, and we present the best results of three of them in the figure. The shaded part of the left figure represents the fluctuation of the reward value in each episode, and the solid line represents the most stable training result. The loss value on the right figure corresponds to the situation when the reward value is the most stable.}
  \label{fig:R&L}
  \vspace{-0.3cm}
\end{figure}

\textbf{Total rewards.} As shown in Fig.~\ref{fig:R&L}(a), at the beginning of training, the total reward value obtained by the agent is relatively low, and the loss function has not yet converged. With the continuous learning of the agent, when the training reaches 150,000 rounds, the reward function values for four algorithms all show a trend of convergence. Although the introduction of the GAIL mechanism does not speed up the convergence of the reward function of the A3C and DPPO models, the total reward value of the A3C-GAIL and DPP-GAIL are higher than the original DRL model, especially the total reward value of the A3C-GAIL is 243\% higher than DeepExploit.

\textbf{Loss value.} In the course of many experiments, we found that the loss value of the benchmark DeepExploit will explode in the later stage of training, which causes the model to fail to convergence. Therefore, we modify the loss function in the original DeepExploit algorithm and use modified DeepExploit as a new benchmark to compare with the other three algorithms. Since the multi-thread training mechanism only calculates the loss value of the model when passing the parameters between the thread and the chief network, the meaning of the ordinate is not equal to the number of training episodes. From the convergence of the loss function of the four algorithm models in Fig.~\ref{fig:R&L}(b), we can see that after updating the DRL algorithm model with GAIL mechanism 10,000 times, the fluctuation of the loss value should be stable to the original DRL model, particularly the loss fluctuation contrast between A3C-GAIL and A3C is more evident than DPPO-GAIL and DPPO.

\begin{figure*}[!htp]
\centering
        \includegraphics[width=0.64\linewidth]{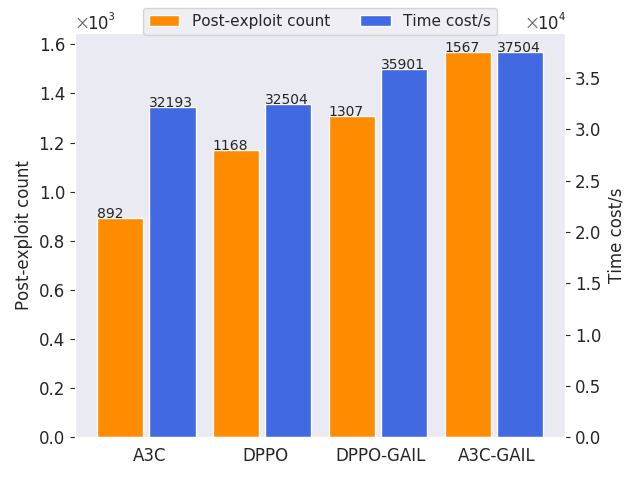}
\caption{Comparison of post-exploit count and time cost of different algorithms in exploiting the target host Metasploitable2 }
\label{fig:20w}
\end{figure*}

\textbf{Post-exploit count \& Time cost.} In addition to the analysis of the total reward and loss value from the perspective of the deep reinforcement learning model, we also compare the penetration performance from the perspective of PT itself with two indicators: the count of successful post-exploit and the time cost for completing 200,000 episodes training of PT. As shown in Fig.~\ref{fig:20w}, after replacing the new and old policies and changing the thread and chief network parameter update mechanism, the post-exploit count of the DPPO model is 276 more than that of DeepExploit, but the time cost only increased by 311 seconds, which indicates that applying DPPO to automate PT is effective and its penetration performance is better than the DeepExploit.

For A3C-GAIL and DPPO-GAIL, we combined the GAIL network then employed the A3C-GAIL to perform intelligent PT. The post-exploit counts of A3C-GAIL are the highest among the four algorithms models. However, the time cost of A3C-GAIL is also the highest, with only 260 times more than the DPPO-GAIL in the count of post-exploit; the time cost is twice as high as DPPO-GAIL, which explains the reason why the total reward value of A3C-GAIL in Fig.~\ref{fig:20w} is higher than that of DPPO-GAIL. Since we only counted the number of successful post-exploit and gave positive rewards for the exploit that did not establish a session with the target host, A3C-GAIL may have performed multiple exploitations in extra time but failed in the post-exploit. Therefore, considering the count of post-exploit and time cost, we can conclude that DPPO-GAIL performs more efficient intelligent PT.

\begin{figure*}[!htp]
\centering
        \includegraphics[width=0.64\linewidth]{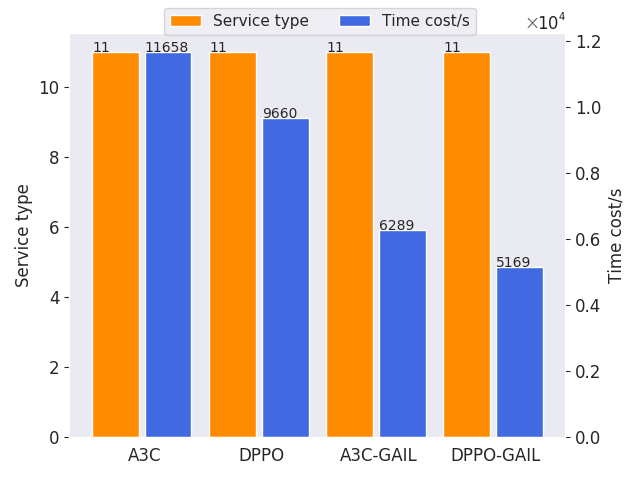}
\caption{Comparison of and time cost of exploiting 11 different types of service vulnerabilities in the target host Meatsploitable2}
\label{fig:pt11}
\end{figure*}

\textbf{Time cost under exploiting limited vulnerabilities.} In order to verify the better performance of applying the DPPO-GAIL model to perform intelligent PT, we conducted another set of comparative experiments. Since Metasploit can only exploit the vulnerabilities of 11 service types in the target host Metasploitable2, we counted the time it takes to exploit these 11 different vulnerabilities for the four algorithms respectively as the last evaluation indicator. Intuitively, from Fig.~\ref{fig:pt11} we can see that under the guidance of expert samples and a combination of the GAIL network, the time cost of the DPPO-GAIL is the least, followed by A3C-GAIL. What is more, the time cost of DPPO-GAIL, A3C-GAIL and DPPO is reduced by 55.7\%, 46.1\%, and 17.1\%  compared with DeepExploit, respectively.

Considering the above five different evaluation indicators adequately, under the guidance of expert sample knowledge and combination with the GAIL network, the performance of A3C-GAIL and DPPO-GAIL in intelligently exploiting the actual target Metasploitable2 has improved in total reward and the count of successful post-exploit compared with DeepExploit. The total reward value has increased by 253\% and 160\%, respectively, and the count of post-exploit has increased by 75.7\% and 46.5\%. However, the time cost of A3C-GAIL was twice as high as that of DPPO-GAIL. Besides, after exploiting 11 different types of service vulnerabilities, the time cost of the two algorithms was reduced by 46.1\% and 55.7\%, respectively.

\vspace{0.1cm}

\vspace{-0.4cm}
%%%%%%%%%%%%%%%%%%%%%%%%%%%%%%%%%%%%%%%% Answer2
\begin{center}
\fcolorbox{black}{white!20}{\parbox{0.97\linewidth}
    {
        \emph{\textbf{Answer to RQ2}}:
        The penetration performance of A3C-GAIL and DPPO-GAIL is better than that of A3C and DPPO. Besides, the DPPO-GAIL can execute more post-exploits successfully, and exploit more vulnerabilities of different service types in less time to achieve the SOTA performance.
    }
}
\end{center}

\subsection{Network scenario}

In order to verify the effectiveness of automating PT in different scale networks based on GAIL, we conducted another part of experiments on three different network scenarios: a small-scale network with or without honeypot, and a large-scale network in a high-fidelity network simulator Nasim. The primary purpose of this scenario is to generate an optimal penetration path in a trial and error training manner, through exploiting one host in one subnet and gaining control of it, then using the compromised host as a stepping stone according to network connection relationships to carry out several lateral movements for obtaining the control of the final sensitive host.

\subsubsection{Experiment settings for different network scenarios}
\begin{figure*}[!htp]
\centering
        \includegraphics[width=0.68\linewidth]{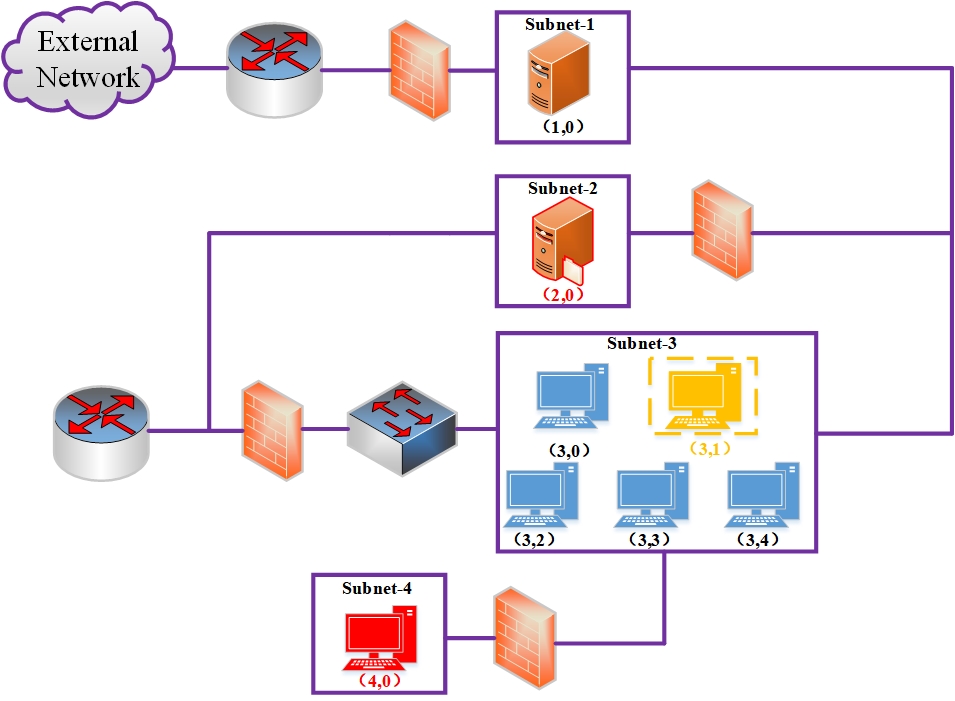}
        \includegraphics[width=0.31\linewidth]{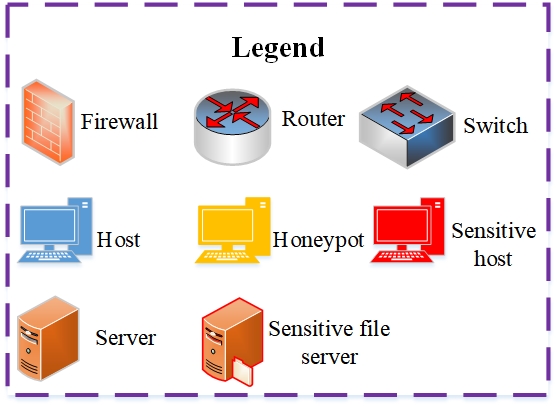}
 \caption{It is the layout of the small-scale network. A honeypot is located in (3,2). In the scene without a honeypot, the location of (3,2) is regarded as a normal host.}
\label{fig:smallscale}
\end{figure*}

At first, we automate PT in a small-scale network (without honeypot) scenario as shown in Fig.~\ref{fig:smallscale}. The network contains 4 sub-networks and 8 hosts, sensitive hosts are (2, 0) and (4, 0), which is the target host in PT. Firewalls block communication between subnets with restricted services, and penetrating from one subnet to another requires a specific cost each time. 

The hosts in the subnet can communicate with each other. The configuration information of each host demonstrating in Table~\ref{tabel:Configuration list}, including the hosts' respective virtual address, operating system, service information, process information, and the value of the host. In order to simulate the action of an actual attacker, we assume that the agent cannot directly obtain network topology information and host configuration information. Therefore, in addition to processing exploits(Exploit) and privilege promotion (Promotion) actions, scanning (Scan) can also be used to obtain relevant information about the host and target network.

For each host, the agent can choose the actions shown in Table~\ref{tabel:action list}. We choose processes and services that attackers often use during the PT process in a subnet to replace network security vulnerabilities, set the probabilities of `0.9' and `0.6' for SSH, HTTP and FTP according to CVSS score, respectively, and set the probability to `1' for privilege promotion and scanning. For example, the agent can get the user permission of the host through exploiting the vulnerability with FTP service then performing the Daclsvc privilege promotion action to get the root permission of the host, thereby realizing a penetration attack on the host.

\begin{table}[!htp]
\footnotesize
\centering
\caption{Configuration list}
\label{tabel:Configuration list}
\resizebox{0.6\linewidth}{!}{
\begin{tabular}{lllll}
\toprule \hline
\textbf{Address} & \textbf{Operation System} &\textbf{Host-value} &\textbf{Service} &\textbf{Process} \\
\hline
(1,0)  & Linux    & 0    & HTTP    & Tomcat   \\
(2,0)  & Linux    & 100    & SSH,HTTP    & /  \\
(3,0)  & Windows    & 0    & HTTP       & /   \\
(3,1)  & Windows    & 0    & FTP,HTTP  & Daclsvc   \\
(3,2)  & Windows    & 0    & FTP,HTTP    & Daclsvc   \\
(3,3)  & Windows    & 0    & FTP    & /  \\
(3,4)  & Windows    & 0    & FTP    & Daclsvc   \\
(4,0)  & Linux    & 100    & SSH,HTTP    & Tomcat   \\
\hline \bottomrule
\end{tabular}
}
\end{table}

Then, we conduct the same experiments in the small-scale network with a honeypot scenario shown in Fig.~\ref{fig:smallscale} as well, optimizing strategy for finding the penetration path through RL algorithm to realize an intelligent PT process. The honeypot is located in (3,2). But in the samll-scale without a honeypot, the location of (3,2) is regarded as a normal host.  We set the value of the honeypot host to -100, assuming the agent knows the address of the honeypot in the subnet. In this case, the agent can bypass the honeypot for exploiting.

\begin{table}[!htp]
\footnotesize
\centering
\caption{Agent action list}
\label{tabel:action list}
\resizebox{0.6\linewidth}{!}{
\begin{tabular}{llllll}
\toprule \hline
\textbf{Name} & \textbf{Type} &\textbf{Operation System} &\textbf{Cost} &\textbf{Prob}  &\textbf{Access}\\ 
\hline
SSH-Exp   & Exploit    & Linux    & 3    & 0.9  &User \\
FTP-Exp  & Exploit    & Windows    & 1    &0.6  &User  \\
HTTP-Exp  & Exploit    & None      & 2   & 0.9  &User   \\
Tomcat  & Promotion    & Linux       & 1     & 1  &Root \\
Daclsvc  & Promotion    & Windows    & 1    & 1   &Root\\
Service-Scan  & Scan    & /    & 1    & 1  &/\\
Os-Scan       & Scan     & /   & 1   & 1  &/\\
Subnet-Scan  & Scan     & /    & 1    & 1  &/\\
Process-Scan  & Scan     & /   & 1    & 1   &/\\
\hline \bottomrule
\end{tabular}
}
\end{table}
At last, we conducted the same PT experiment in a large-scale simulated network scenario with 23 hosts. Different from the small-scale network scenario, the large-scale network scenario is more complicated: as shown in Fig.~\ref{fig:largescale}, it has 8 subnets and a total of 23 hosts. There is a sensitive host in subnet 2 and subnet 7, respectively (2, 0) and (7, 0), which are the target hosts of the PT. The hosts in the remaining subnets are all normal hosts; each host runs various services and communicates. Besides, only subnet 1 could communicate with the external network directly. The remaining 6 subnets communicated in the internal network; subnet 5 could only communicate with subnet 3 directly, and subnet 6 could only communicate with subnet 4 directly. For example, if the attacker arrives at (5,0), his goal is to exploit the sensitive host (7,0). Therefore, he has to come back to subnet 3 for lateral movements and other operations to exploit one host in subnet 4, then arrive at (7.0) through the same operation. It significantly increases the cost of the attack, so our purpose is to find the optimal penetration path at the lowest cost.

Besides, there are three operating systems types: Windows, Linux and Unix; seven different exploiting services: SSH, FTP, HTTP, RPC, PHP, Samba, SMTP (Simple Mail Transfer Protocol), and three privilege promotion process: Tomcat, Daclsvc, and Schtask in the large-scale network scenario. We set the probabilities of `0.9' for SSH, HTTP; set `0.6' for FTP, RPC, PHP and SMTP; set `0.3' for Samba according to CVSS score respectively, and set the probability to `1' for privilege promotion and scanning. The penetration mechanism of this scene is similar to the small-scale network.

However, due to the increase of exploiting services and hosts, the input state in the large-scale scene is expanded to 768 dimensions, and also the output action is expanded to 322 dimensions, which is a high-dimensional input and output scene. Due to the higher complexity of large-scale network scenarios, the combination of reinforcement learning to optimize the penetration path has the problem of too large state dimensions and sparse reward values. Therefore, we also introduce the expert sample knowledge into this scene and combine the GAIL network with training the agent for finding and optimizing the penetration path.
\begin{figure*}[!htp]
\centering
        \includegraphics[width=0.77\linewidth]{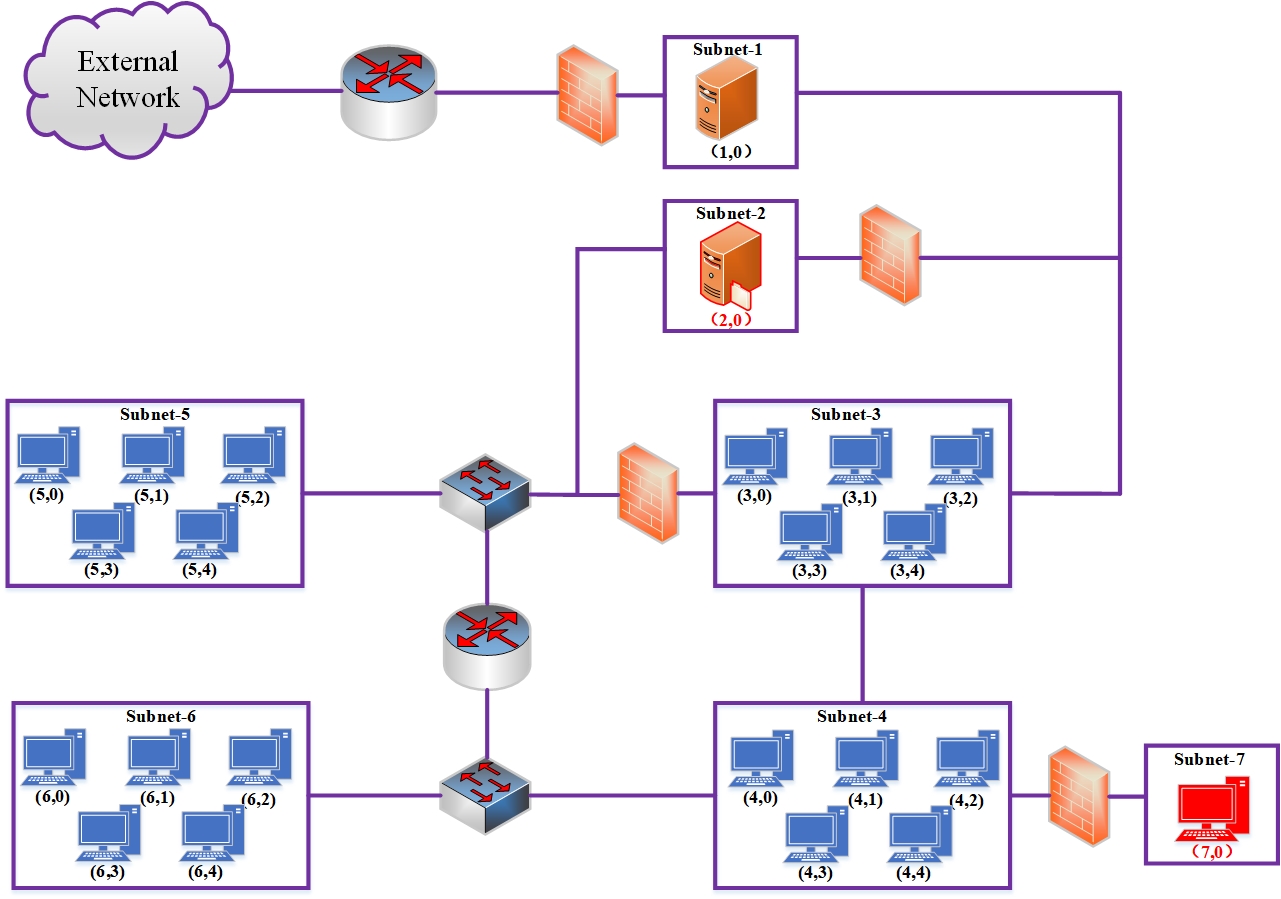}
\caption{It is the layout of the large-scale network. Sensitive file server and sensitive host located in (2,0) and (7,0), respectively.}
\label{fig:largescale}
\end{figure*}

The reward value of successfully exploiting the sensitive host is 100 in the small scale network with or without the honeypot scenario illustrated in Fig.~\ref{fig:smallscale}, but the reward value of falling into the honeypot is set to -100. 
Likewise, in the large-scale network shown in Fig.~\ref{fig:largescale}, setting the reward values of successfully exploiting each of two sensitive hosts to 100. Agents can only execute exploit between connected subnets or hosts in the network. The ending of each round of the agent are the following two situations:
\begin{itemize}
\item Obtaining root permissions for all sensitive hosts.
\item Training steps in each round have reached the maximum value.
\end{itemize}

The agent is tried to obtain the maximum cumulative reward value through continuous trial and error, thereby learning the optimal penetration path for exploiting the target sensitive host. In the small-scale with a honeypot network, since the reward value of falling into the honeypot is negative, the agent will choose to bypass the honeypot during the PT process and finally learn the optimal penetration path to exploit the sensitive host with fewer steps.

We first use the Q-learning algorithm to conduct experiments and then perform automated PT on the small-scale, small-scale with honeypot and large-scale networks. Second, expert knowledge is introduced based on the benchmark algorithm Q-learning and combines the GAIL network with Q-learning. At last, compare the penetration performance of Q-learning-GAIL and Q-learning in finding the optimal penetration path. Besides, we set the training round is 20000 for each scenario, recorded the steps and cost needed to generate the penetration path in each round, and adopted the average value as the experiment results. For two RL algorithms in three different scenarios: Q-learning and Q-learning-GAIL, we apply them to automate PT and evaluate their performance. \subsubsection{Evaluation metrics}
In the training phase, we calculating the average value every 1000 rounds, and applying the following three indicators to analyze the answers of \emph{\textbf{RQ3}} and \emph{\textbf{RQ4}}:
\begin{itemize}
\item \textbf{Average reward.} In the training process, the average round reward is calculated every certain round, which can be used to evaluate the overall performance of the algorithm;
\item \textbf{Average steps.} In the training process, the average round length is calculated every certain round, which represents the time cost required for the generation of the penetration path, and is also used to evaluate the overall performance of the algorithm.
\item \textbf{Probability of invading honeypot.} In the process of PT, the average probability of honeypot intrusion is calculated every certain round to evaluate the effectiveness of honeypot deployment.
\end{itemize}

\subsubsection{Research question 3}
\begin{center}
\fcolorbox{black}{gray!20}{\parbox{0.97\linewidth}
    {
       Does expanding the expert sample amount still improve the penetration performance in different network scenarios?
    }
}
\end{center}

We will answer the question through analyzing the experiment results from the one indicator: Training rounds.

The optimal path is definite when optimizing the penetration path in a network scene. Therefore, there is no situation where a particular state corresponds to multiple optimal penetration paths. In other words, when the agent finds the best penetration path, the state and action at this time are a one-to-one relationship. What is more, there is no concept of the complexity of expert knowledge rules. Therefore, we are only considering the impact of expert sample amount on penetration performance.

We evaluated the impact of different expert sample amounts on the improvement of penetration performance in three different network scenarios separately and used the training rounds required to find the optimal penetration path as the indicator. It can be seen from Table~\ref{tabel:virtualamonut} that the sample size of expert knowledge has a specific impact on the training effect of optimizing the penetration path.
Without introducing expert sample knowledge, the training rounds required to find the optimal penetration path in the small-scale network, small-scale with honeypot network scenario, and large-scale network scenario are 4000, 5000, and 3000, respectively. When the sample amount expands to 5000 pairs, the rounds for generating the optimal penetration path are 3000, 3500, and 2000, respectively, reduced by 1000, 1500, and 1000 rounds in turn. 

On this basis, the expansion of the expert knowledge sample size did not significantly accelerate generating the optimal penetration path. Therefore, we all introduced 5000 pairs of expert knowledge sample amounts in the following three scene experiments.

\begin{table}[!htp]
\footnotesize
\centering
\caption{Training rounds for different expert sample amount  to find the  \\optimal penetration path in different network scenarios}
\label{tabel:virtualamonut}
\resizebox{0.70\linewidth}{!}{
\begin{tabular}{lllll}
\toprule \hline
\textbf{\diagbox{Network scene}{Expert sample amount}} & \textbf{0} &\textbf{2500} &\textbf{5000} &\textbf{7500}  \\ 
\hline
Small-scale   & 4000    & 3500   & \textbf{3000}    & 3000  \\
Small-scale with honeypot  & 5000    & {4000}    & \textbf{3500}    &3500  \\
Large-scale  & 3000    & 3000     & \textbf{2000}   & 2000     \\
\hline \bottomrule
\end{tabular}
}
\end{table}

\vspace{-0.3cm}
%%%%%%%%%%%%%%%%%%%%%%%%%%%%%%%%%%%%%%%% Answer2
\begin{center}
\fcolorbox{black}{white!20}{\parbox{0.97\linewidth}
    {
        \emph{\textbf{Answer to RQ3}}:
          Penetration performance improved with expanding the expert samples in the range of 5,000 pairs, but it is not significantly improved when the expert sample exceeds this range.
    }
}
\end{center}

\subsubsection{Research question 4}
\begin{center}
\fcolorbox{black}{gray!20}{\parbox{0.97\linewidth}
    {
       Can GAIL also achieve the SOTA penetration performance in different network scenarios?
    }
}
\end{center}

We will answer this question from three different network scenarios combined with the three indicators: Average reward, Average step and Probability of invading honeypot through the following experimental results.

\begin{figure}[t]
\centering
\subfigure[Comparison of average reward]{
    \includegraphics[width=0.48\linewidth]{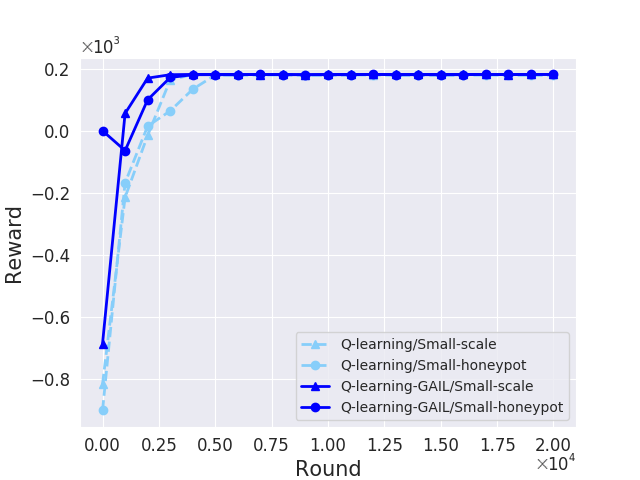}
    }
\subfigure[Comparison of average steps]{
    \includegraphics[width=0.48\linewidth]{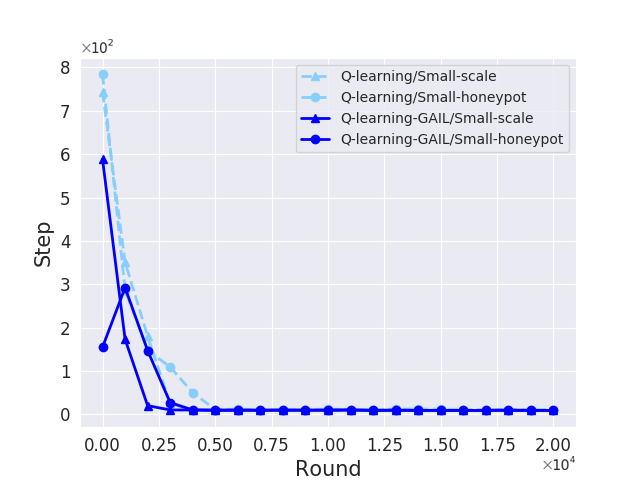}
    }
    \vspace{-0.4cm}
  \caption{Comparison of average reward (left) and average steps (right) in small-scale network with or without honeypot scenarios by Q-learning and Q-learning-GAIL with 20000 rounds training.}
  \label{fig:5}
  \vspace{-0.3cm}
  \label{fig:smallandhoney}
\end{figure}

\textbf{Small-scale network} 
It can be seen from the comparison of the average reward value and average steps in Fig.~\ref{fig:smallandhoney}, with the introduction of expert knowledge samples and the combination of the GAIL mechanism, accelerate the convergence of the reward function and training steps, and finally improve the penetration performance to a certain extent. In a small-scale network scenario without a honeypot, using Q-learning-GAIL algorithm to optimize the penetration path, the average reward value began to converge in 2000 rounds, and the average training steps also showed the same trend, both of them are about 1000 rounds faster than the Q-learning. 

As to small-scale with a honeypot, the average reward value of each round is higher than Q-learning. The fluctuation of the reward value before convergence is much gentle than that of Q-learning. In terms of training steps, the average round steps of the Q-learning-GAIL algorithm tend to converge at 3000 rounds, while Q-learning tends to converge at 5000 rounds. The average reward value also shows this trend, and both are about 2000 rounds faster than Q-learning for finding the optimal penetration path.

In short, the introduction of expert knowledge samples speeds up finding the optimal penetration path in the small-scale network. Because the complexity of the small network is relatively lower than large-scale, the introduction of the GAIL mechanism does not reduce the average round steps and the steps of finding optimal penetration stabilizing at 9 steps in the end. On the other side, training costs are also reduced. Especially in the honeypot scenario, the reduction in training cost is twice that in the case of no honeypot.

\begin{figure*}[!htp]
\centering
        \includegraphics[width=0.5\linewidth]{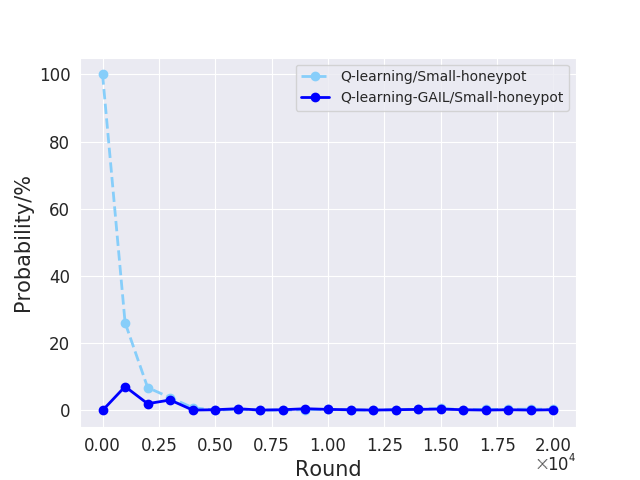}
\caption{Probability comparison of invading the honeypot  }
\label{fig:pro}
\end{figure*}
\textbf{Small-scale network with honeypot} In addition, this section comparative analyzes the intrusion probability of honeypots in the small-scale network with a honeypot. Intuitively, this indicator reflects the effectiveness of deploying honeypots, that is, whether the agent will fall into the honeypot again after training through expert knowledge and GAIL network guidance. The agent's learning process aims to exploit sensitive hosts, not to invade the honeypot. We set invading the honeypot (3, 2) a negative reward. Therefore, the lower the probability of falling into the honeypot, the better the expert sample guides the agent to train, which also indicates that the agent's action distribution is infinitely close to the expert sample at this time.

Fig.~\ref{fig:pro} intuitively reflects that the probability that the agent trained by Q-learning-GAIL will fall into the honeypot is much lower than the benchmark Q-learning in the first 3000 rounds. At the beginning of training, the probability of falling into the honeypot is 100\% training by Q-learning,  while training with Q-learning-GAIL is directly reduced to 0. Although in the next 3000 rounds, the probability that GAIL guides the agent to fall into a honeypot fluctuates slightly, in the entire 20000 rounds, the probability of applying Q-learning-GAIL algorithm to fall into honeypot is lower than Q-learning in each round, which also indicates that the introduction of the GAIL mechanism improves the penetration performance more significantly in the network scenarios with honeypot.

\textbf{Large-scale network} Compared with the small-scale network, it is not difficult to find that, when applying Q-learning-GAIL to determine optimal penetration path, the improvement of PT performance is more evident in the large-scale network, as shown in 
Fig.~\ref{fig:largescale}.

In the early stage of training, the agent is still in the exploring stage. The average reward and average steps of the benchmark Q-learning in the first 3000 rounds show a major fluctuations trend. However, under the guidance of expert knowledge, the Q-learning-GAIL algorithm shows a relatively stable performance in the early stage of the training agent for optimal penetration path-finding. The average reward value is positive in the first 3000 rounds, and the average exploration steps are within 50 steps. After 2000 rounds, both the average reward value and the average steps show a convergence trend. Moreover, the final average steps stabilized at 15 steps, while the final average steps of Q-learning is 17-18 steps, which 2-3 steps have accelerated in the final round.
\begin{figure}[t]
\centering
\subfigure[Comparison of average reward]{
    \includegraphics[width=0.48\linewidth]{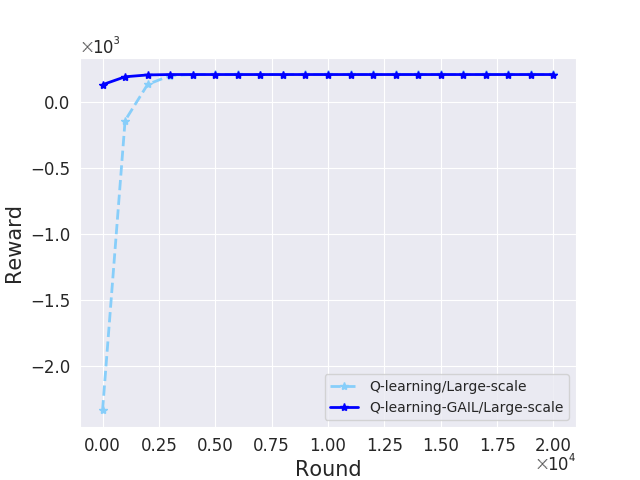}
    }
\subfigure[Comparison of average steps]{
    \includegraphics[width=0.48\linewidth]{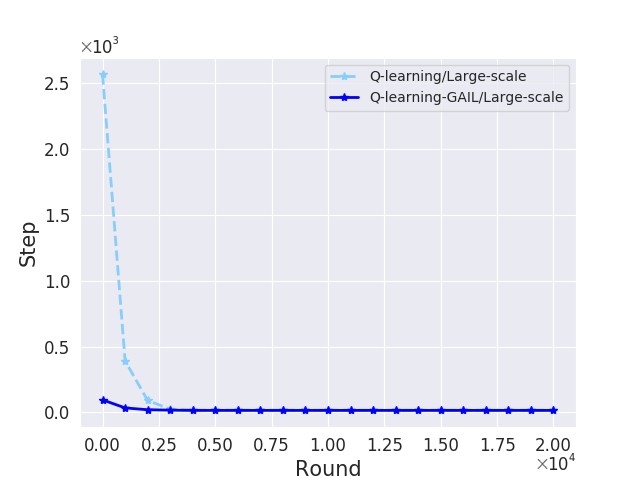}
    }
    \vspace{-0.4cm}
  \caption{Comparison of average reward (left) and average steps (right) in large-scale network scenarios}
  \label{fig:5}
  \vspace{-0.3cm}
  \label{fig:largescale}
\end{figure}
In short, by introducing expert knowledge and combining the GAIL network with Q-learning to optimize penetration path, Q-learning-GAIL shows a more stable penetration performance and achieves the SOTA penetration performance with less time and cost.

\vspace{-0.3cm}
%%%%%%%%%%%%%%%%%%%%%%%%%%%%%%%%%%%%%%%% Answer4
\begin{center}
\fcolorbox{black}{white!20}{\parbox{0.97\linewidth}
    {
        \emph{\textbf{Answer to RQ4}}:
         Whether in a small-scale network with honeypot or not or in a large-scale network, the penetration performance of Q-learning-GAIL has all reached the SOTA. Especially in small-scale with honeypot and large-scale networks, penetration performance improved even more apparent.
    }
}
\end{center}

\section{Conclusion and future work\label{Conclusion and future work}}
This paper proposed a method to automate the penetration testing process based on GAIL, and evaluated the effectiveness and performance of exploiting the single host and three different networks. The imitation learning agent is modeled as a penetration attacker in the actual and simulated scenarios. In the experiments, we compared the methods of A3C-GAIL, DPPO-GAIL and Q-learning-GAIL, which are our methods, with the DeepExploit, DPPO and Q-learning in exploiting single host and network. The results testified the performance of our methods.In addition to combining with A3C, DPPO and Q-learning, GAIL can also combine with other DRL / RL algorithms. What is more, our original intention is to propose a general penetration testing framework GAIL-PT that can help conjunct GAIL with different DRL / RL algorithms to automate the penetration testing process in different scenarios. To our best knowledge, it is the first study of applying GAIL for penetration testing.  We publish the code on Github so that interested scholars can conduct related research. 

In aspect of limitation of the proposed method, its complexity of A3C-GAIL and DPPO-GAIL is relatively high. Although the network simulator NASim appears accurate vulnerability service and specific penetration operations in network scenarios, there are still differences between NASim and the existing network, and our method has not yet been validated in real complex networks. Therefore, future work can be carried out in the following two aspects: reducing the algorithm's complexity and applying GAIL-PT to exploit the real complex network.

\section{Acknowledgment}
This research was supported by the National Natural Science Foundation of China (No. 62072406), the National Key Laboratory of Science and Technology on Information System Security (No. 61421110502), the National Key R\&D Projects of China (No. 2018AAA0100801), 
the Key R\&D Projects in Zhejiang Province (No. 2021C01117), the 2020 Industrial Internet Innovation Development Project (No.TC200H01V), and ``Ten Thousand Talents Program" in Zhejiang Province (No. 2020R52011).

% \section*{References}


\begin{thebibliography}{58}
\expandafter\ifx\csname natexlab\endcsname\relax\def\natexlab#1{#1}\fi
\providecommand{\url}[1]{\texttt{#1}}
\providecommand{\href}[2]{#2}
\providecommand{\path}[1]{#1}
\providecommand{\DOIprefix}{doi:}
\providecommand{\ArXivprefix}{arXiv:}
\providecommand{\URLprefix}{URL: }
\providecommand{\Pubmedprefix}{pmid:}
\providecommand{\doi}[1]{\href{http://dx.doi.org/#1}{\path{#1}}}
\providecommand{\Pubmed}[1]{\href{pmid:#1}{\path{#1}}}
\providecommand{\bibinfo}[2]{#2}
\ifx\xfnm\relax \def\xfnm[#1]{\unskip,\space#1}\fi
%Type = Article
\bibitem[{Arce and McGraw(2004)}]{2}
\bibinfo{author}{Arce, I.}, \bibinfo{author}{McGraw, G.}, \bibinfo{year}{2004}.
\newblock \bibinfo{title}{Guest editors' introduction: Why attacking systems is
  a good idea}.
\newblock \bibinfo{journal}{IEEE Security \& Privacy} \bibinfo{volume}{2},
  \bibinfo{pages}{17--19}.
%Type = Article
\bibitem[{Arkin et~al.(2005)Arkin, Stender and McGraw}]{1}
\bibinfo{author}{Arkin, B.}, \bibinfo{author}{Stender, S.},
  \bibinfo{author}{McGraw, G.}, \bibinfo{year}{2005}.
\newblock \bibinfo{title}{Software penetration testing}.
\newblock \bibinfo{journal}{IEEE Security Privacy} \bibinfo{volume}{3},
  \bibinfo{pages}{84--87}.
\newblock \DOIprefix\doi{10.1109/MSP.2005.23}.
%Type = Article
\bibitem[{Arora and Doshi(2021)}]{im11}
\bibinfo{author}{Arora, S.}, \bibinfo{author}{Doshi, P.}, \bibinfo{year}{2021}.
\newblock \bibinfo{title}{A survey of inverse reinforcement learning:
  Challenges, methods and progress}.
\newblock \bibinfo{journal}{Artificial Intelligence} , \bibinfo{pages}{103500}.
%Type = Article
\bibitem[{Berner et~al.(2019)Berner, Brockman, Chan, Cheung, D{\k{e}}biak,
  Dennison, Farhi, Fischer, Hashme, Hesse et~al.}]{openai5}
\bibinfo{author}{Berner, C.}, \bibinfo{author}{Brockman, G.},
  \bibinfo{author}{Chan, B.}, \bibinfo{author}{Cheung, V.},
  \bibinfo{author}{D{{e}}biak, P.}, \bibinfo{author}{Dennison, C.},
  \bibinfo{author}{Farhi, D.}, \bibinfo{author}{Fischer, Q.},
  \bibinfo{author}{Hashme, S.}, \bibinfo{author}{Hesse, C.}, et~al.,
  \bibinfo{year}{2019}.
\newblock \bibinfo{title}{Dota 2 with large scale deep reinforcement learning}.
\newblock \bibinfo{journal}{arXiv preprint arXiv:1912.06680} .
%Type = Article
\bibitem[{Bland et~al.(2020)Bland, Petty, Whitaker, Maxwell and
  Cantrell}]{bland2020}
\bibinfo{author}{Bland, J.A.}, \bibinfo{author}{Petty, M.D.},
  \bibinfo{author}{Whitaker, T.S.}, \bibinfo{author}{Maxwell, K.P.},
  \bibinfo{author}{Cantrell, W.A.}, \bibinfo{year}{2020}.
\newblock \bibinfo{title}{Machine learning cyberattack and defense strategies}.
\newblock \bibinfo{journal}{Computers \& security} \bibinfo{volume}{92},
  \bibinfo{pages}{101738}.
%Type = Article
\bibitem[{Canese et~al.(2021)Canese, Cardarilli, Di~Nunzio, Fazzolari,
  Giardino, Re and Span{\`o}}]{trialanderror}
\bibinfo{author}{Canese, L.}, \bibinfo{author}{Cardarilli, G.C.},
  \bibinfo{author}{Di~Nunzio, L.}, \bibinfo{author}{Fazzolari, R.},
  \bibinfo{author}{Giardino, D.}, \bibinfo{author}{Re, M.},
  \bibinfo{author}{Span{\`o}, S.}, \bibinfo{year}{2021}.
\newblock \bibinfo{title}{Multi-agent reinforcement learning: A review of
  challenges and applications}.
\newblock \bibinfo{journal}{Applied Sciences} \bibinfo{volume}{11},
  \bibinfo{pages}{4948}.
%Type = Article
\bibitem[{Dulac-Arnold et~al.(2015)Dulac-Arnold, Evans, van Hasselt, Sunehag,
  Lillicrap, Hunt, Mann, Weber, Degris and Coppin}]{largespace}
\bibinfo{author}{Dulac-Arnold, G.}, \bibinfo{author}{Evans, R.},
  \bibinfo{author}{van Hasselt, H.}, \bibinfo{author}{Sunehag, P.},
  \bibinfo{author}{Lillicrap, T.}, \bibinfo{author}{Hunt, J.},
  \bibinfo{author}{Mann, T.}, \bibinfo{author}{Weber, T.},
  \bibinfo{author}{Degris, T.}, \bibinfo{author}{Coppin, B.},
  \bibinfo{year}{2015}.
\newblock \bibinfo{title}{Deep reinforcement learning in large discrete action
  spaces}.
\newblock \bibinfo{journal}{arXiv preprint arXiv:1512.07679} .
%Type = Article
\bibitem[{El~Kamel et~al.(2020)El~Kamel, Eddabbah, Lmoumen and
  Touahni}]{honeypot2}
\bibinfo{author}{El~Kamel, N.}, \bibinfo{author}{Eddabbah, M.},
  \bibinfo{author}{Lmoumen, Y.}, \bibinfo{author}{Touahni, R.},
  \bibinfo{year}{2020}.
\newblock \bibinfo{title}{A smart agent design for cyber security based on
  honeypot and machine learning}.
\newblock \bibinfo{journal}{Security and Communication Networks}
  \bibinfo{volume}{2020}.
%Type = Inproceedings
\bibitem[{Elderman et~al.(2017)Elderman, Pater, Thie, Drugan and
  Wiering}]{elderman2017}
\bibinfo{author}{Elderman, R.}, \bibinfo{author}{Pater, L.J.},
  \bibinfo{author}{Thie, A.S.}, \bibinfo{author}{Drugan, M.M.},
  \bibinfo{author}{Wiering, M.A.}, \bibinfo{year}{2017}.
\newblock \bibinfo{title}{Adversarial reinforcement learning in a cyber
  security simulation.}, in: \bibinfo{booktitle}{ICAART (2)}, pp.
  \bibinfo{pages}{559--566}.
%Type = Inproceedings
\bibitem[{Fan et~al.(2020)Fan, Wang, Xie and Yang}]{DQN}
\bibinfo{author}{Fan, J.}, \bibinfo{author}{Wang, Z.}, \bibinfo{author}{Xie,
  Y.}, \bibinfo{author}{Yang, Z.}, \bibinfo{year}{2020}.
\newblock \bibinfo{title}{A theoretical analysis of deep q-learning}, in:
  \bibinfo{booktitle}{Learning for Dynamics and Control},
  \bibinfo{organization}{PMLR}. pp. \bibinfo{pages}{486--489}.
%Type = Inproceedings
\bibitem[{Farquhar et~al.(2020)Farquhar, Gustafson, Lin, Whiteson, Usunier and
  Synnaeve}]{growingactionspace}
\bibinfo{author}{Farquhar, G.}, \bibinfo{author}{Gustafson, L.},
  \bibinfo{author}{Lin, Z.}, \bibinfo{author}{Whiteson, S.},
  \bibinfo{author}{Usunier, N.}, \bibinfo{author}{Synnaeve, G.},
  \bibinfo{year}{2020}.
\newblock \bibinfo{title}{Growing action spaces}, in:
  \bibinfo{booktitle}{International Conference on Machine Learning},
  \bibinfo{organization}{PMLR}. pp. \bibinfo{pages}{3040--3051}.
%Type = Article
\bibitem[{Goodfellow et~al.(2014)Goodfellow, Pouget-Abadie, Mirza, Xu,
  Warde-Farley, Ozair, Courville and Bengio}]{GAN}
\bibinfo{author}{Goodfellow, I.}, \bibinfo{author}{Pouget-Abadie, J.},
  \bibinfo{author}{Mirza, M.}, \bibinfo{author}{Xu, B.},
  \bibinfo{author}{Warde-Farley, D.}, \bibinfo{author}{Ozair, S.},
  \bibinfo{author}{Courville, A.}, \bibinfo{author}{Bengio, Y.},
  \bibinfo{year}{2014}.
\newblock \bibinfo{title}{Generative adversarial nets}.
\newblock \bibinfo{journal}{Advances in neural information processing systems}
  \bibinfo{volume}{27}.
%Type = Techreport
\bibitem[{Haeni(1997)}]{firewall1}
\bibinfo{author}{Haeni, R.E.}, \bibinfo{year}{1997}.
\newblock \bibinfo{title}{Firewall penetration testing}.
\newblock \bibinfo{type}{Technical Report}. Citeseer.
%Type = Article
\bibitem[{He et~al.(2016)He, Dai and Ning}]{he2016}
\bibinfo{author}{He, X.}, \bibinfo{author}{Dai, H.}, \bibinfo{author}{Ning,
  P.}, \bibinfo{year}{2016}.
\newblock \bibinfo{title}{Faster learning and adaptation in security games by
  exploiting information asymmetry}.
\newblock \bibinfo{journal}{IEEE Transactions on Signal Processing}
  \bibinfo{volume}{64}, \bibinfo{pages}{3429--3443}.
%Type = Misc
\bibitem[{HelpSysthems(2021)}]{coreimpact}
\bibinfo{author}{HelpSysthems}, \bibinfo{year}{2021}.
\newblock \bibinfo{title}{Core impact}.
\newblock
  \bibinfo{note}{\url{https://www.coresecurity.com/products/core-impact/}}.
%Type = Article
\bibitem[{Ho and Ermon(2016)}]{GAIL}
\bibinfo{author}{Ho, J.}, \bibinfo{author}{Ermon, S.}, \bibinfo{year}{2016}.
\newblock \bibinfo{title}{Generative adversarial imitation learning}.
\newblock \bibinfo{journal}{Advances in neural information processing systems}
  \bibinfo{volume}{29}, \bibinfo{pages}{4565--4573}.
%Type = Article
\bibitem[{{Ho} et~al.(2016){Ho}, {Gupta} and {Ermon}}]{im16}
\bibinfo{author}{{Ho}, J.}, \bibinfo{author}{{Gupta}, J.K.},
  \bibinfo{author}{{Ermon}, S.}, \bibinfo{year}{2016}.
\newblock \bibinfo{title}{{Model-Free Imitation Learning with Policy
  Optimization}}.
\newblock \bibinfo{journal}{arXiv e-prints} ,
  \bibinfo{pages}{arXiv:1605.08478}\href{http://arxiv.org/abs/1605.08478}{\tt
  arXiv:1605.08478}.
%Type = Inproceedings
\bibitem[{Hu et~al.(2020)Hu, Beuran and Tan}]{autoPTDQN}
\bibinfo{author}{Hu, Z.}, \bibinfo{author}{Beuran, R.}, \bibinfo{author}{Tan,
  Y.}, \bibinfo{year}{2020}.
\newblock \bibinfo{title}{Automated penetration testing using deep
  reinforcement learning}, in: \bibinfo{booktitle}{2020 IEEE European Symposium
  on Security and Privacy Workshops (EuroS\&PW)}, \bibinfo{organization}{IEEE}.
  pp. \bibinfo{pages}{2--10}.
%Type = Inproceedings
\bibitem[{Ingols et~al.(2009)Ingols, Chu, Lippmann, Webster and Boyer}]{BID8}
\bibinfo{author}{Ingols, K.}, \bibinfo{author}{Chu, M.},
  \bibinfo{author}{Lippmann, R.}, \bibinfo{author}{Webster, S.},
  \bibinfo{author}{Boyer, S.}, \bibinfo{year}{2009}.
\newblock \bibinfo{title}{Modeling modern network attacks and countermeasures
  using attack graphs}, in: \bibinfo{booktitle}{2009 Annual Computer Security
  Applications Conference}, \bibinfo{organization}{IEEE}. pp.
  \bibinfo{pages}{117--126}.
%Type = Misc
\bibitem[{{Isao Takaesu}(2018)}]{DeepExploit}
\bibinfo{author}{{Isao Takaesu}}, \bibinfo{year}{2018}.
\newblock \bibinfo{title}{Metasploit meets machine learning}.
\newblock \bibinfo{note}{\url{https://www.mbsd.jp/blog/20180228.html}}.
%Type = Article
\bibitem[{Kaur and Kaur(2017)}]{os}
\bibinfo{author}{Kaur, G.}, \bibinfo{author}{Kaur, N.}, \bibinfo{year}{2017}.
\newblock \bibinfo{title}{Penetration testing--reconnaissance with nmap tool.}
\newblock \bibinfo{journal}{International Journal of Advanced Research in
  Computer Science} \bibinfo{volume}{8}.
%Type = Article
\bibitem[{Kaushik and Ojha(2014)}]{ds}
\bibinfo{author}{Kaushik, M.}, \bibinfo{author}{Ojha, G.},
  \bibinfo{year}{2014}.
\newblock \bibinfo{title}{Attack penetration system for sql injection}.
\newblock \bibinfo{journal}{International journal of advanced computer
  research} \bibinfo{volume}{4}, \bibinfo{pages}{724}.
%Type = Book
\bibitem[{Kennedy et~al.(2011)Kennedy, O'gorman, Kearns and Aharoni}]{3}
\bibinfo{author}{Kennedy, D.}, \bibinfo{author}{O'gorman, J.},
  \bibinfo{author}{Kearns, D.}, \bibinfo{author}{Aharoni, M.},
  \bibinfo{year}{2011}.
\newblock \bibinfo{title}{Metasploit: the penetration tester's guide}.
\newblock \bibinfo{publisher}{No Starch Press}.
%Type = Article
\bibitem[{Levine et~al.(2011)Levine, Popovic and Koltun}]{im15}
\bibinfo{author}{Levine, S.}, \bibinfo{author}{Popovic, Z.},
  \bibinfo{author}{Koltun, V.}, \bibinfo{year}{2011}.
\newblock \bibinfo{title}{Nonlinear inverse reinforcement learning with
  gaussian processes}.
\newblock \bibinfo{journal}{Advances in neural information processing systems}
  \bibinfo{volume}{24}, \bibinfo{pages}{19--27}.
%Type = Misc
\bibitem[{Mapper(2021)}]{nmap}
\bibinfo{author}{Mapper, N.}, \bibinfo{year}{2021}.
\newblock \bibinfo{title}{Nmap}.
\newblock \bibinfo{note}{\url{https://nmap.org/}}.
%Type = Inproceedings
\bibitem[{McDaniel et~al.(2016)McDaniel, Talvi and Hay}]{CTF}
\bibinfo{author}{McDaniel, L.}, \bibinfo{author}{Talvi, E.},
  \bibinfo{author}{Hay, B.}, \bibinfo{year}{2016}.
\newblock \bibinfo{title}{Capture the flag as cyber security introduction}, in:
  \bibinfo{booktitle}{2016 49th hawaii international conference on system
  sciences (hicss)}, \bibinfo{organization}{IEEE}. pp.
  \bibinfo{pages}{5479--5486}.
%Type = Inproceedings
\bibitem[{Mell et~al.(2007)Mell, Scarfone, Romanosky et~al.}]{cvss}
\bibinfo{author}{Mell, P.}, \bibinfo{author}{Scarfone, K.},
  \bibinfo{author}{Romanosky, S.}, et~al., \bibinfo{year}{2007}.
\newblock \bibinfo{title}{A complete guide to the common vulnerability scoring
  system version 2.0}, in: \bibinfo{booktitle}{Published by FIRST-forum of
  incident response and security teams}, p.~\bibinfo{pages}{23}.
%Type = Inproceedings
\bibitem[{Mnih et~al.(2016)Mnih, Badia, Mirza, Graves, Lillicrap, Harley,
  Silver and Kavukcuoglu}]{ACA3C}
\bibinfo{author}{Mnih, V.}, \bibinfo{author}{Badia, A.P.},
  \bibinfo{author}{Mirza, M.}, \bibinfo{author}{Graves, A.},
  \bibinfo{author}{Lillicrap, T.}, \bibinfo{author}{Harley, T.},
  \bibinfo{author}{Silver, D.}, \bibinfo{author}{Kavukcuoglu, K.},
  \bibinfo{year}{2016}.
\newblock \bibinfo{title}{Asynchronous methods for deep reinforcement
  learning}, in: \bibinfo{booktitle}{International conference on machine
  learning}, \bibinfo{organization}{PMLR}. pp. \bibinfo{pages}{1928--1937}.
%Type = Article
\bibitem[{Mnih et~al.(2013)Mnih, Kavukcuoglu, Silver, Graves, Antonoglou,
  Wierstra and Riedmiller}]{DRL}
\bibinfo{author}{Mnih, V.}, \bibinfo{author}{Kavukcuoglu, K.},
  \bibinfo{author}{Silver, D.}, \bibinfo{author}{Graves, A.},
  \bibinfo{author}{Antonoglou, I.}, \bibinfo{author}{Wierstra, D.},
  \bibinfo{author}{Riedmiller, M.}, \bibinfo{year}{2013}.
\newblock \bibinfo{title}{Playing atari with deep reinforcement learning}.
\newblock \bibinfo{journal}{arXiv preprint arXiv:1312.5602} .
%Type = Article
\bibitem[{Moyer and Schultz(1996)}]{firewall2}
\bibinfo{author}{Moyer, P.R.}, \bibinfo{author}{Schultz, E.E.},
  \bibinfo{year}{1996}.
\newblock \bibinfo{title}{A systematic methodology for firewall penetration
  testing}.
\newblock \bibinfo{journal}{Network Security} \bibinfo{volume}{1996},
  \bibinfo{pages}{11--18}.
%Type = Inproceedings
\bibitem[{Neal et~al.(2021)Neal, Dagdougui, Lodi and Fernandez}]{4}
\bibinfo{author}{Neal, C.}, \bibinfo{author}{Dagdougui, H.},
  \bibinfo{author}{Lodi, A.}, \bibinfo{author}{Fernandez, J.M.},
  \bibinfo{year}{2021}.
\newblock \bibinfo{title}{Reinforcement learning based penetration testing of a
  microgrid control algorithm}, in: \bibinfo{booktitle}{2021 IEEE 11th Annual
  Computing and Communication Workshop and Conference (CCWC)},
  \bibinfo{organization}{IEEE}. pp. \bibinfo{pages}{0038--0044}.
%Type = Inproceedings
\bibitem[{Phillips and Swiler(1998)}]{graph}
\bibinfo{author}{Phillips, C.}, \bibinfo{author}{Swiler, L.P.},
  \bibinfo{year}{1998}.
\newblock \bibinfo{title}{A graph-based system for network-vulnerability
  analysis}, in: \bibinfo{booktitle}{Proceedings of the 1998 workshop on New
  security paradigms}, pp. \bibinfo{pages}{71--79}.
%Type = Inproceedings
\bibitem[{Pozdniakov et~al.(2020)Pozdniakov, Alonso, Stankovic, Tam and
  Jones}]{aduit}
\bibinfo{author}{Pozdniakov, K.}, \bibinfo{author}{Alonso, E.},
  \bibinfo{author}{Stankovic, V.}, \bibinfo{author}{Tam, K.},
  \bibinfo{author}{Jones, K.}, \bibinfo{year}{2020}.
\newblock \bibinfo{title}{Smart security audit: Reinforcement learning with a
  deep neural network approximator}, in: \bibinfo{booktitle}{2020 International
  Conference on Cyber Situational Awareness, Data Analytics and Assessment
  (CyberSA)}, \bibinfo{organization}{IEEE}. pp. \bibinfo{pages}{1--8}.
%Type = Inproceedings
\bibitem[{Qiu et~al.(2014a)Qiu, Jia, Wang, Xia and Lv}]{BID9}
\bibinfo{author}{Qiu, X.}, \bibinfo{author}{Jia, Q.}, \bibinfo{author}{Wang,
  S.}, \bibinfo{author}{Xia, C.}, \bibinfo{author}{Lv, L.},
  \bibinfo{year}{2014}a.
\newblock \bibinfo{title}{Automatic generation algorithm of penetration graph
  in penetration testing}, in: \bibinfo{booktitle}{2014 Ninth International
  Conference on P2P, Parallel, Grid, Cloud and Internet Computing},
  \bibinfo{organization}{IEEE}. pp. \bibinfo{pages}{531--537}.
%Type = Inproceedings
\bibitem[{Qiu et~al.(2014b)Qiu, Wang, Jia, Xia and Xia}]{BID6}
\bibinfo{author}{Qiu, X.}, \bibinfo{author}{Wang, S.}, \bibinfo{author}{Jia,
  Q.}, \bibinfo{author}{Xia, C.}, \bibinfo{author}{Xia, Q.},
  \bibinfo{year}{2014}b.
\newblock \bibinfo{title}{An automated method of penetration testing}, in:
  \bibinfo{booktitle}{2014 IEEE Computers, Communications and IT Applications
  Conference}, \bibinfo{organization}{IEEE}. pp. \bibinfo{pages}{211--216}.
%Type = Misc
\bibitem[{{RAPID7}(2017)}]{metasploitable2}
\bibinfo{author}{{RAPID7}}, \bibinfo{year}{2017}.
\newblock \bibinfo{title}{Metasploitable2}.
\newblock
  \bibinfo{note}{\url{https://docs.rapid7.com/metasploit/metasploitable-2/Download:https://docs.rapid7.com/metasploit/metasploitable-2/}}.
%Type = Misc
\bibitem[{RAPID7(2019)}]{nexpose}
\bibinfo{author}{RAPID7}, \bibinfo{year}{2019}.
\newblock \bibinfo{title}{Nexpose}.
\newblock \bibinfo{note}{\url{https://www.rapid7.com/products/nexpose/}}.
%Type = Misc
\bibitem[{RAPID7(2021)}]{metasploit}
\bibinfo{author}{RAPID7}, \bibinfo{year}{2021}.
\newblock \bibinfo{title}{Metasploit}.
\newblock \bibinfo{note}{\url{http://www.metasploit.com/}}.
%Type = Inproceedings
\bibitem[{Sarraute et~al.(2012)Sarraute, Buffet and Hoffmann}]{POMDP2}
\bibinfo{author}{Sarraute, C.}, \bibinfo{author}{Buffet, O.},
  \bibinfo{author}{Hoffmann, J.}, \bibinfo{year}{2012}.
\newblock \bibinfo{title}{Pomdps make better hackers: Accounting for
  uncertainty in penetration testing}, in: \bibinfo{booktitle}{Twenty-Sixth
  AAAI Conference on Artificial Intelligence}.
%Type = Article
\bibitem[{Sarraute et~al.(2013)Sarraute, Buffet and Hoffmann}]{POMDP1}
\bibinfo{author}{Sarraute, C.}, \bibinfo{author}{Buffet, O.},
  \bibinfo{author}{Hoffmann, J.}, \bibinfo{year}{2013}.
\newblock \bibinfo{title}{Penetration testing== pomdp solving?}
\newblock \bibinfo{journal}{arXiv preprint arXiv:1306.4714} .
%Type = Article
\bibitem[{Schulman et~al.(2017)Schulman, Wolski, Dhariwal, Radford and
  Klimov}]{snoop33}
\bibinfo{author}{Schulman, J.}, \bibinfo{author}{Wolski, F.},
  \bibinfo{author}{Dhariwal, P.}, \bibinfo{author}{Radford, A.},
  \bibinfo{author}{Klimov, O.}, \bibinfo{year}{2017}.
\newblock \bibinfo{title}{Proximal policy optimization algorithms}.
\newblock \bibinfo{journal}{arXiv preprint arXiv:1707.06347} .
%Type = Article
\bibitem[{Schwartz and Kurniawati(2019)}]{explode}
\bibinfo{author}{Schwartz, J.}, \bibinfo{author}{Kurniawati, H.},
  \bibinfo{year}{2019}.
\newblock \bibinfo{title}{Autonomous penetration testing using reinforcement
  learning}.
\newblock \bibinfo{journal}{arXiv preprint arXiv:1905.05965} .
%Type = Inproceedings
\bibitem[{Schwartz et~al.(2020)Schwartz, Kurniawati and El-Mahassni}]{POMDP3}
\bibinfo{author}{Schwartz, J.}, \bibinfo{author}{Kurniawati, H.},
  \bibinfo{author}{El-Mahassni, E.}, \bibinfo{year}{2020}.
\newblock \bibinfo{title}{Pomdp+ information-decay: Incorporating defender's
  behaviour in autonomous penetration testing}, in:
  \bibinfo{booktitle}{Proceedings of the International Conference on Automated
  Planning and Scheduling}, pp. \bibinfo{pages}{235--243}.
%Type = Misc
\bibitem[{Schwartz and Kurniawatti(2019)}]{nasim}
\bibinfo{author}{Schwartz, J.}, \bibinfo{author}{Kurniawatti, H.},
  \bibinfo{year}{2019}.
\newblock \bibinfo{title}{Nasim: Network attack simulator}.
\newblock \bibinfo{note}{\url{https://networkattacksimulator.readthedocs.io/}}.
%Type = Article
\bibitem[{Silver et~al.(2016a)Silver, Huang, Maddison, Guez, Sifre, Van
  Den~Driessche, Schrittwieser, Antonoglou, Panneershelvam, Lanctot
  et~al.}]{Alphago}
\bibinfo{author}{Silver, D.}, \bibinfo{author}{Huang, A.},
  \bibinfo{author}{Maddison, C.J.}, \bibinfo{author}{Guez, A.},
  \bibinfo{author}{Sifre, L.}, \bibinfo{author}{Van Den~Driessche, G.},
  \bibinfo{author}{Schrittwieser, J.}, \bibinfo{author}{Antonoglou, I.},
  \bibinfo{author}{Panneershelvam, V.}, \bibinfo{author}{Lanctot, M.}, et~al.,
  \bibinfo{year}{2016}a.
\newblock \bibinfo{title}{Mastering the game of go with deep neural networks
  and tree search}.
\newblock \bibinfo{journal}{nature} \bibinfo{volume}{529},
  \bibinfo{pages}{484--489}.
%Type = Article
\bibitem[{Silver et~al.(2016b)Silver, Huang, Maddison, Guez, Sifre, Van
  Den~Driessche, Schrittwieser, Antonoglou, Panneershelvam, Lanctot
  et~al.}]{im1}
\bibinfo{author}{Silver, D.}, \bibinfo{author}{Huang, A.},
  \bibinfo{author}{Maddison, C.J.}, \bibinfo{author}{Guez, A.},
  \bibinfo{author}{Sifre, L.}, \bibinfo{author}{Van Den~Driessche, G.},
  \bibinfo{author}{Schrittwieser, J.}, \bibinfo{author}{Antonoglou, I.},
  \bibinfo{author}{Panneershelvam, V.}, \bibinfo{author}{Lanctot, M.}, et~al.,
  \bibinfo{year}{2016}b.
\newblock \bibinfo{title}{Mastering the game of go with deep neural networks
  and tree search}.
\newblock \bibinfo{journal}{nature} \bibinfo{volume}{529},
  \bibinfo{pages}{484--489}.
%Type = Book
\bibitem[{Spitzner(2003)}]{honeypot1}
\bibinfo{author}{Spitzner, L.}, \bibinfo{year}{2003}.
\newblock \bibinfo{title}{Honeypots: tracking hackers}.
  volume~\bibinfo{volume}{1}.
\newblock \bibinfo{publisher}{Addison-Wesley Reading}.
%Type = Article
\bibitem[{Sun et~al.(2020)Sun, Li, Zhang and Liu}]{im10}
\bibinfo{author}{Sun, R.}, \bibinfo{author}{Li, X.}, \bibinfo{author}{Zhang,
  L.}, \bibinfo{author}{Liu, J.}, \bibinfo{year}{2020}.
\newblock \bibinfo{title}{Distributed storage codes based on double-layered
  piggybacking framework}.
\newblock \bibinfo{journal}{IEEE Access} \bibinfo{volume}{8},
  \bibinfo{pages}{150447--150464}.
%Type = Misc
\bibitem[{Tenale(2021)}]{nessus}
\bibinfo{author}{Tenale}, \bibinfo{year}{2021}.
\newblock \bibinfo{title}{Nessus}.
\newblock \bibinfo{note}{\url{https://zh-cn.tenable.com/products/nessus}}.
%Type = Article
\bibitem[{Torabi et~al.(2018)Torabi, Warnell and Stone}]{im9}
\bibinfo{author}{Torabi, F.}, \bibinfo{author}{Warnell, G.},
  \bibinfo{author}{Stone, P.}, \bibinfo{year}{2018}.
\newblock \bibinfo{title}{Behavioral cloning from observation}.
\newblock \bibinfo{journal}{arXiv preprint arXiv:1805.01954} .
%Type = Article
\bibitem[{Tran et~al.(2021)Tran, Akella, Standen, Kim, Bowman, Richer and
  Lin}]{HA-DRL}
\bibinfo{author}{Tran, K.}, \bibinfo{author}{Akella, A.},
  \bibinfo{author}{Standen, M.}, \bibinfo{author}{Kim, J.},
  \bibinfo{author}{Bowman, D.}, \bibinfo{author}{Richer, T.},
  \bibinfo{author}{Lin, C.T.}, \bibinfo{year}{2021}.
\newblock \bibinfo{title}{Deep hierarchical reinforcement agents for automated
  penetration testing}.
\newblock \bibinfo{journal}{arXiv preprint arXiv:2109.06449} .
%Type = Article
\bibitem[{Vinyals et~al.(2019)Vinyals, Babuschkin, Czarnecki, Mathieu, Dudzik,
  Chung, Choi, Powell, Ewalds, Georgiev et~al.}]{Alphastar}
\bibinfo{author}{Vinyals, O.}, \bibinfo{author}{Babuschkin, I.},
  \bibinfo{author}{Czarnecki, W.M.}, \bibinfo{author}{Mathieu, M.},
  \bibinfo{author}{Dudzik, A.}, \bibinfo{author}{Chung, J.},
  \bibinfo{author}{Choi, D.H.}, \bibinfo{author}{Powell, R.},
  \bibinfo{author}{Ewalds, T.}, \bibinfo{author}{Georgiev, P.}, et~al.,
  \bibinfo{year}{2019}.
\newblock \bibinfo{title}{Grandmaster level in starcraft ii using multi-agent
  reinforcement learning}.
\newblock \bibinfo{journal}{Nature} \bibinfo{volume}{575},
  \bibinfo{pages}{350--354}.
%Type = Inproceedings
\bibitem[{Wang et~al.(2019)Wang, Ciliberto, Amadori and Demiris}]{im13}
\bibinfo{author}{Wang, R.}, \bibinfo{author}{Ciliberto, C.},
  \bibinfo{author}{Amadori, P.V.}, \bibinfo{author}{Demiris, Y.},
  \bibinfo{year}{2019}.
\newblock \bibinfo{title}{Random expert distillation: Imitation learning via
  expert policy support estimation}, in: \bibinfo{booktitle}{International
  Conference on Machine Learning}, \bibinfo{organization}{PMLR}. pp.
  \bibinfo{pages}{6536--6544}.
%Type = Article
\bibitem[{Wiering and Van~Otterlo(2012)}]{RL}
\bibinfo{author}{Wiering, M.A.}, \bibinfo{author}{Van~Otterlo, M.},
  \bibinfo{year}{2012}.
\newblock \bibinfo{title}{Reinforcement learning}.
\newblock \bibinfo{journal}{Adaptation, learning, and optimization}
  \bibinfo{volume}{12}.
%Type = Article
\bibitem[{Zennaro and Erdodi(2020)}]{modelptql}
\bibinfo{author}{Zennaro, F.M.}, \bibinfo{author}{Erdodi, L.},
  \bibinfo{year}{2020}.
\newblock \bibinfo{title}{Modeling penetration testing with reinforcement
  learning using capture-the-flag challenges and tabular q-learning}.
\newblock \bibinfo{journal}{arXiv preprint arXiv:2005.12632} .
%Type = Article
\bibitem[{ZHANG et~al.(2020)ZHANG, Tianyang, Junhu and Qingxian}]{pathfind}
\bibinfo{author}{ZHANG, Y.}, \bibinfo{author}{Tianyang, Z.},
  \bibinfo{author}{Junhu, Z.}, \bibinfo{author}{Qingxian, W.},
  \bibinfo{year}{2020}.
\newblock \bibinfo{title}{Domain-independent intelligent planning technology
  and its application to automated penetration testing oriented attack path
  discovery}.
\newblock \bibinfo{journal}{Journal of Electronics \& Information Technology}
  \bibinfo{volume}{42}, \bibinfo{pages}{2095--2107}.
%Type = Article
\bibitem[{Zhou et~al.(2021)Zhou, Liu, Hou, Zhong and Zhang}]{zhouimDQN}
\bibinfo{author}{Zhou, S.}, \bibinfo{author}{Liu, J.}, \bibinfo{author}{Hou,
  D.}, \bibinfo{author}{Zhong, X.}, \bibinfo{author}{Zhang, Y.},
  \bibinfo{year}{2021}.
\newblock \bibinfo{title}{Autonomous penetration testing based on improved deep
  q-network}.
\newblock \bibinfo{journal}{Applied Sciences} \bibinfo{volume}{11},
  \bibinfo{pages}{8823}.
%Type = Inproceedings
\bibitem[{Ziebart et~al.(2008)Ziebart, Maas, Bagnell, Dey et~al.}]{maxentropy}
\bibinfo{author}{Ziebart, B.D.}, \bibinfo{author}{Maas, A.L.},
  \bibinfo{author}{Bagnell, J.A.}, \bibinfo{author}{Dey, A.K.}, et~al.,
  \bibinfo{year}{2008}.
\newblock \bibinfo{title}{Maximum entropy inverse reinforcement learning.}, in:
  \bibinfo{booktitle}{Aaai}, \bibinfo{organization}{Chicago, IL, USA}. pp.
  \bibinfo{pages}{1433--1438}.

\end{thebibliography}
\end{document}